\documentclass[a4paper,12pt]{elsarticle}

\usepackage[top=0.5in,bottom=0.75in,left=0.6in,right=0.6in]{geometry}
\usepackage{graphicx}

\usepackage[figurename=Fig.]{caption}
\usepackage[tablename=Tab.]{caption}
\usepackage{color}
\usepackage{float}
\usepackage{amssymb}
\usepackage{amsmath}
\usepackage{mathrsfs}
\usepackage{textcomp}
\usepackage{bbm}
\usepackage{cases}
\usepackage{upgreek}
\usepackage{makecell}
\usepackage{pdflscape}
\usepackage{subfigure}
\usepackage{pdfpages}

\usepackage{makeidx}
\usepackage{bm}
\usepackage{xstring}

\usepackage{epstopdf}
\usepackage{lipsum}
\usepackage{slashed}
\usepackage{multicol}
\usepackage{setspace}
\usepackage{booktabs}
\usepackage[table]{xcolor}
\usepackage{multirow}
\usepackage{braket}

\usepackage[separate-uncertainty]{siunitx}
\usepackage[displaymath, mathlines]{lineno}
\newcommand{\eref}[1]{Eq.\,(\ref{#1})}
\newcommand{\w}{\omega}
\newcommand{\vo}{\vec{o}\@ifnextchar{^}{\,}{}}
\def\up{\mathrm}

\renewcommand{\thetable}{\Roman{table}}

\graphicspath{{PDF/}}

\def\hm{\hspace{-0.1em}}
\def\hp{\hspace{0.1em}}
\def\slashchar#1{\setbox0=\hbox{$#1$}           % set a box for #1
   \dimen0=\wd0                                 % and get its size
   \setbox1=\hbox{/} \dimen1=\wd1               % get size of /
   \ifdim\dimen0>\dimen1                        % #1 is bigger
      \rlap{\hbox to \dimen0{\hfil/\hfil}}      % so center / in box
      #1                                        % and print #1
   \else                                        % / is bigger
      \rlap{\hbox to \dimen1{\hfil$#1$\hfil}}   % so center #1
      /                                         % and print /
   \fi}                                         %
\def\sl#1{\setbox0=\hbox{#1}
  \dimen0=\wd0
  \rlap{\hbox to \dimen0{\hss/\hss}}%
  #1}

\definecolor{myblue}{RGB}{65,111,166}
\definecolor{myred}{RGB}{168,66,63}
\definecolor{mygreen}{RGB}{134,164,74}
\definecolor{mypurple}{RGB}{110,84,141}
\definecolor{myindigo}{RGB}{61,150,174}
\definecolor{myorange}{RGB}{218,129,55}
\definecolor{mylightblue}{RGB}{142,165,203}

\makeatletter
\def\ps@pprintTitle{%
  \let\@oddhead\@empty
  \let\@evenhead\@empty
  \let\@oddfoot\@empty
  \let\@evenfoot\@oddfoot
}
\makeatother



\usepackage{hyperref}
\hypersetup{
    colorlinks=true,
    linkcolor=red,
    citecolor=red,
    urlcolor=cyan,
}
\begin{document}

\begin{frontmatter}
\title{Mass spectra and wave functions for the doubly heavy baryons with $J^P=1^+$ heavy diquark cores}

\author[1,2]{Qiang Li}\ead{liruo@nwpu.edu.cn}
\author[2,3,4]{Chao-Hsi Chang}\ead{zhangzx@itp.ac.cn}
\author[5]{Si-Xue Qin}\ead{sqin@cqu.edu.cn}
%\author[1]{Xu-Chang Zheng}\ead{zhengxc@itp.ac.cn}
\author[6]{Guo-Li Wang}\ead{gl\_wang@hit.edu.cn}
%\cortext[corauthor]{Corresponding author}

\address[1]{Department of Applied Physics, Northwestern Polytechnical University, Xi'an 710072, China}
\address[2]{Institute of Theoretical Physics, Chinese Academy of Sciences, Beijing 100190, China}
\address[3]{School of Physical Sciences, University of Chinese Academy of Sciences, 19A Yuquan Road, Beijing 100049, China}
\address[4]{CCAST (World Laboratory), P.O. Box 8730, Beijing 100190, China}
\address[5]{Department of Physics, Chongqing University, Chongqing 401331, China}
\address[6]{Department of Physics, Hebei University, Baoding 071002, China}

\begin{abstract}
Mass spectra and wave functions for the doubly heavy baryons are computed in the picture that the two heavy quarks inside a baryon form a compact heavy `diquark core' in color anti-triplet, then bind the rest light quark into a colorless baryon. The reduced two two-body problems are described by the relativistic Bethe-Salpeter equations (BSEs) with relevant QCD-inspired kernels. So far, in this work, we only focus on the doubly heavy baryons with $1^+$ heavy diquark cores. By solving the BSEs in the instantaneous approximation, we first present mass spectra and the relativistic wave functions for diquark cores, and then those for the low-lying states of $J^P=\frac{1}{2}^+$ and $\frac{3}{2}^+$ baryons with flavors $(ccq)$, $(bcq)$, and $(bbq)$. Comparisons with other approaches are also presented.


\end{abstract}

\end{frontmatter}

\section{Introduction}
The LHCb collaboration reported their observation on the doubly charmed baryon $\Xi_{cc}^{++}(ccu)$ in the $\Lambda^+_cK^-\pi^+\pi^+$ decay channel recently, where the mass and lifetime of the baryon are determined as $3.621$ GeV\,\cite{LHCb2017B} and 0.256 ps\,\cite{LHCb2018}, and thus systematically theoretical researches on the kindred baryons (doubly heavy baryons) become imperative. In addition, LHCb also reported observations of five narrow $\Omega_c^0$ excited baryons in 2017\,\cite{LHCb2017A} and $\Xi^-_{b}$ in 2018\,\cite{LHCb2018-Xib}. In most theoretical models, the mass of the doubly heavy baryon $\Xi_{cc}^{+(+)}$ is predicted in the range $3.5\sim3.7$ GeV \cite{Chernyshev1996,Gershtein1998,Gershtein1999,Gershtein2000,Ebert2002,HeDH2004,Chang2006,ZhangJR2008,WangZG2010,Brodsky2011,Aliev2012,SunZF2015,WeiKW2015,SunZF2016}. The mass splitting between $\Xi_{cc}^{++}$ and $\Xi^+_{cc}$ is predicted to be several MeV due to the mass difference of the light quarks $u$ and $d$. The predicted mass by lattice QCD is about 3.6 GeV\,\cite{Lewis2001,Flynn2003,Brown2014}, which is quite close to the LHCb's observation. The lifetimes of $\Xi^{+}_{cc}$ and $\Xi_{cc}^{++}$ are predicted to be quite long as $50\sim250$ and $200\sim700$ fs\,\cite{Karliner2014,Kiselev1999,Chang2008,Berezhnoy2016}, respectively. The LHCb's measurement on $\Xi_{cc}^{++}$ is very close to the lower boundary of the theoretical predictions. It is expected that more information on the doubly heavy baryon(s) will be avaiable in the near future.

Compared with mesons, baryons as three-quark bound states are much more complicated. However, for a doubly heavy baryon, the involved two heavy quarks move relatively slow, and thus are bound with each other strongly to form a color anti-triplet diquark core\footnote{According to QCD and color confinement, the gluon exchange interaction between two quarks is attractive only when they are in a color anti-triplet. Moreover only in a color anti-triplet the diquark can combine with a third quark to form a colorless baryon. Thus throughout this work the diquark cores are meant to be in a color anti-triplet, i.e. the color part of the diquark-core wave function is always anti-symmetric.}.  Thus,  it is reasonable to reduce the three-body problem into two two-body ones.
%The QCD-inspired interaction, e.g. one gluon exchange being dominant, between the two heavy quarks is attractive if they are in a color anti-triplet, but is repulsive if they are in a color sextet state\footnote{Two quarks may be in two possible color states only: anti-symmetric one in color anti-triplet or symmetric one in color sextet.}. The two heavy quarks being bound into a heavy diquark core needs an attractive interaction, and only being in color anti-triplet, the heavy diquark core may form a colorless doubly heavy baryon when it combines one more light quark, thus here the considered heavy diquark core must be in a color anti-triplet state. Later on without precise statement, the diquark core would always mean it in color anti-triplet.
According to the Pauli principle, the baryon wave functions must be totally anti-symmetric under the interchange of any two quarks. For the diquark core with even orbital angle momentum $L_\up{D}$, the spin-flavor part of the wave function must be symmetric. Hence, $(cc)$- and $(bb)$-diquarks at ground states ($S$-wave) must have $J^P=1^+$, while $(bc)$-diquark, carrying different flavors, may have $J^P=1^+$ or $0^+$. %Note here $j$ is used to denote the diquark total angular momentum and $J$ will be used to denote the baryon total angular momentum. 
So far, in this work, we only focus on the $\frac{1}{2}^+$ or $\frac{3}{2}^+$ baryons with $1^+$ doubly heavy diquark cores ($cc$, $bc$ or $bb$). Whereas the $\frac{1}{2}^+$ baryons with $0^+$ $(bc)$-diquark, e.g., $\Xi'_{bc}$ and $\Omega'_{bc}$, will be presented elsewhere.

Based on the analysis above, we reduce the three-body problem for the doubly heavy baryons into two two-body problems. First, we deal with diquark cores in the framework of Bethe-Salpeter equation (BSE) with a QCD-inspired kernel. In the so-called instantaneous approximation, we solve the corresponding three-dimensional (Bethe-)Salpeter equation, and obtain the diquark spectra and wave functions. Then, the second step is to establish the BSE for the diquark core and the remaining light quark, where the structured diquark effects in the BSE kernel are described by the corresponding form factors. For solving the baryon BSE, the instantaneous approximation is also adopted. In this scheme, both steps only deal with two-body (Bethe-)Salpeter equations. In fact, Refs.\,\cite{Cahill1987,Keiner1996,Keiner1996A,Maris2002,Maris2004,Maris2005,GuoXH1999,GuoXH2007,WengMH2011,ZhangL2013,LiuY2015,WeiKW2017,LiuLL2017,YUQX2018} have adopted the similar strategy with different approximations or methods. It is known that the BSE framework has been successfully used to study the problems of two-body meson systems, e.g., mass spectra\cite{Chang2005A,Chang2010}, hadronic transitions, electro-weak decays, and etc\,\cite{Chang2005,WangZ2012A,WangT2013,WangT2013A,LiQ2016,LiQ2017,LiQ2017A}. The universal agreement between the theoretical predictions and experimental observations is achieved, so here we would like to push the BSE approach further to the present baryon study. Namely, based on BSE to develop a precise and systematic approach to describe the doubly heavy baryons is our main motivation for this work.

This paper is organized as follows. In section \ref{Sec-2}, we describe the $1^+$ diquark BSE in the instantaneous approximation, solve the mass spectra and wave functions, and then compute the diquark form factors with the Mandelstam formulation. In section \ref{Sec-3}, we derive the three-dimensional diquark-quark (Bethe-)Salpeter equation, and solve the mass spectra and wave functions for baryons with $J^P=\frac{1}{2}^+$ and $\frac{3}{2}^+$. Especially, the diquark-core form factors play a quite important role in determining the BS kernel for the baryons.  In section \ref{Sec-4}, we discuss the obtained results and then compare them with those from other sources. Finally, we provide a brief summary of this work.

\section{Bethe-Salpeter equation for a heavy diquark core and the relevant form factors}\label{Sec-2}
\subsection{\textup{Interaction kernel}}

Since the QCD-inspired interaction kernel for doubly heavy diquark cores has the same root as that of doubly heavy mesons, we briefly review the latter first. In this work, it is assumed that the instantaneous approximation (IA) works well since the considered systems always involve the heavy quark. In the IA, the kernel has no dependence on the time-component of exchanged momentum, and can be expressed as a `revised one-gluon exchange' form as
\begin{equation}
iK_\up{M}(q\,)\simeq i V_\up{M}(\vec q\,)\gamma^\alpha \otimes \gamma_\alpha=i\left[V_\up{G}(\vec q\,)+V_\up{S}(\vec q\,)\right]\gamma^\alpha \otimes \gamma_\alpha,
\end{equation}
where $V_\up{G}(\vec q\,)$ is the usual one-gluon exchange potential and $V_\up{S}(\vec q\,)$ is a phenomenological screened potential, they behave specifically in the Coulomb gauge  as\,\cite{Kim2004}:
\begin{equation}
V_\up{G}(\vec q\,)=-\frac{4}{3} \frac{4\pi \alpha_s(\vec q\,)}{\vec q\,^2+a_1^2},~~~~V_\up{S}(\vec q\,)=\left[(2\pi)^3 \delta^3(\vec q\,)\left( \frac{\lambda}{a_2}+V_0 \right)- \frac{8\pi \lambda}{(\vec q\,^2+a_2^2)^2}\right],
\end{equation}
where $\frac{4}{3}$ is the color factor; $a_{1(2)}$ is introduced to avoid the divergence in the region of small transfer momenta; the screened potential $V_\up{S}(\vec q\,)$ is phenomenologically introduced and characterized by the string constant $\lambda$ and the factor $a_2$. %^\footnote{With small $a_2$, if translating into coordinate space, it is easy to see that $V_\up{C}$ behaves just as the requested confinement effect, `linear increasing' as  the radial increasing in the concerned `long distance' space region.}. 
The potential herein originates from the famous Cornell potential\,\cite{Eichten1978,Eichten1980}, which expresses the one-gluon exchange Coulomb-type potential at short distance and a linear growth confinement one at long distance. In order to incorporate the screening effects\,\cite{Laermann1986,Born1989}, the confinement potential is modified as the aforementioned form\,\cite{Chao1992,DingYB1993,DingYB1995}, where $V_0$ is a free constant fixed by fitting data. Note that a screened potential containing time-like vector type has been widely used and discussed in many works\,\cite{Munz1994,Resag1995,Parramore1995,Zoller1995,Babutsidze1998}, and could achieve well description on the meson systems. Here we assumed $V_\up{S}$ also arises from the gluon exchange and hence carries the same color structure with $V_\up{G}$. The strong coupling constant $\alpha_s$ has the form,
\begin{equation}\notag
\alpha_s(\vec q\,)=\frac{12\pi}{(33-2N_f)}\frac{1}{\ln\left(a+\frac{\vec q\,^2}{\Lambda^2_{\up{QCD}}}\right)},
\end{equation}
where $\Lambda_\up{QCD}$ is the scale of the strong interaction, $N_f$ is the active flavor number, and $a=e$ is a constant regulator.  For later convenience, we split $V_\up{M}(\vec q\,)$ into two parts,
\begin{equation}
V_\up{M}(\vec q\,)%=V_\up{Coul}(\vec q\,)+V_\up{Conf}(\vec q\,)
=(2\pi)^3\delta^3(\vec q\,)V_\up{M1}+V_\up{M2}(\vec q\,),
\end{equation}
where $V_\up{M1}$ is a constant and $V_\up{M2}$ contains all dependence on $\vec q$, i.e.,
\begin{equation}
V_\up{M1} \equiv  \frac{\lambda}{a_2}+V_0 ,~~~V_\up{M2}(\vec q\,)\equiv - \frac{8\pi
\lambda}{(\vec q\,^2+a_2^2)^2}-\frac{4}{3} \frac{4\pi \alpha_s(\vec q\,)}{\vec q\,^2+a_1^2}.
\end{equation}

To construct the BSE kernel for the diquark, we need to transform an anti-fermion line into a fermion one by the charge conjugation. Considering the fact that the quark-antiquark pair in a meson is a color singlet, while the quark-quark pair inside a baryon is a color anti-triplet, and thus the corresponding color factors are $\frac{4}{3}$ and $-\frac{2}{3}$, respectively, hence the kernel for the diquark core is simply assumed to be,
\[
iK_\up{D}(\vec q\,)=-\frac{1}{2} iV_\up{M}(\vec q\,)\gamma^\alpha \otimes (\gamma_\alpha)^\up{T}.
\]
This `half rule' used here is widely adopted in previous works involved the baryons and quark-quark bound problems\,\cite{Gershtein2000,Ebert2002,Karliner2014}. It is exact in the one-gluon exchange limit and has been found to work well in the measurement of the $\Xi_{cc}^{++}$ mass. Its successful extension beyond weak coupling implies that the heavy quark potential factorizes into a color dependent and a space-dependent part, with the latter being the same
for quark–quark and quark–antiquark pairs. The relative factor of $\frac{1}{2}$
then results from the color algebra, just as in the weak-coupling limit\,\cite{Karliner2017}.
Also note that, for heavy quarks, compared with the temporal component ($\alpha=0$) of the vector-vector $\gamma^\alpha \otimes \gamma_\alpha$ interaction kernel, the spatial components ($\alpha=1,2,3$) are suppressed by a factor $v\sim \frac{|\vec{p}\hspace{0.05em}|}{M}$ and will be ignored in numerical calculations. This is also consistent with the analysis in Ref.\,\cite{Olsson1995}.

\subsection{\textup{Bethe-Salpeter equation for the heavy diquark cores}}

\begin{figure}[h!]
\vspace{0.5em}
\centering
\subfigure[BSE of the meson.]{\includegraphics[width=0.45\textwidth]{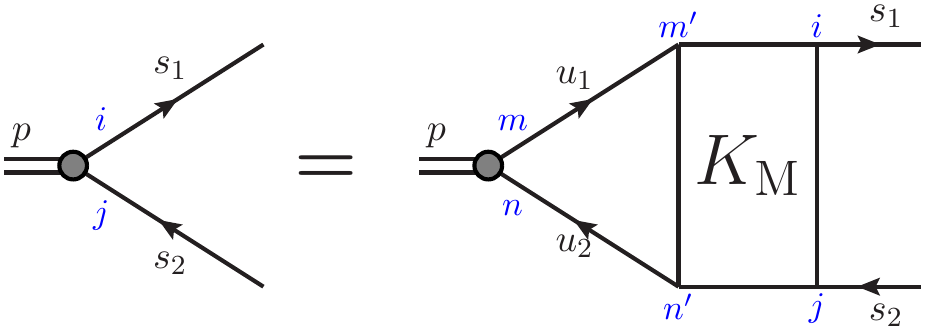} \label{Fig-BS-M}}
~~~~
\subfigure[BSE of the diquark.]{\includegraphics[width=0.45\textwidth]{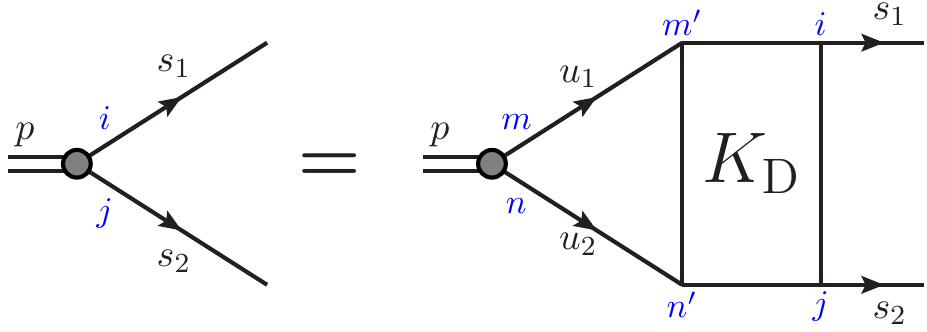} \label{Fig-BS-D}}\\
\caption{BSE of a meson and a diquark. $p$ denotes the bound state momentum, and $p^2=\mu^2$, where $\mu$ is the bound system mass; $s_{1(2)}$ and $u_{1(2)}$ are the quark (antiquark) momenta; and the blue Romans denote the corresponding Dirac indices.}\label{Fig-BSE}
\vspace{0.5em}
\end{figure}
First, we briefly review the Bethe-Salpeter equation for mesons, as shown in \autoref{Fig-BS-M}, which 
in the momentum space reads
\begin{equation}\label{E-BS-Meson}
 \begin{gathered}
\Gamma_\up{M}(p,s)=\int \frac{\up{d}^4 u}{(2\pi)^4} iK_\up{M}(s-u) [S(s_1)\Gamma_\up{M}(p,s)S(-s_2)],
\end{gathered}
\end{equation}
where $\Gamma_\up{M}$ denotes the BS vertex; $S(s_{1})$ and $S(-s_2)$ are the propagators of the quark and antiquark, respectively; the corresponding BS wave function is defined as $\psi_\up{M}(p,s) \equiv S(s_1)\Gamma_\up{M}(p,s)S(-s_2)$. The quark internal momenta $s$ and $u$ are defined respectively as
\[s=\lambda_2s_1-\lambda_1s_2,~~~u=\lambda_2 u_1-\lambda_1 u_2,\]
where $\lambda_i \equiv \frac{\mu_i}{\mu_1+\mu_2}~(i=1,2)$ and $\mu_i$ is the constituent quark mass. The BSE normalization condition can be generally expressed as
\begin{equation}
\begin{gathered}
-i\int \int \frac{\up{d}^4 s}{(2\pi)^4}\frac{\up{d}^4 u}{(2\pi)^4} \up{Tr} ~ \bar{\psi}_\up{M}(p,s) \frac{\partial}{\partial p^0} \left[ I(p,s,u) \right] \psi_\up{M}(p,u) =2p^0,\\
I(p,s,u)= S^{-1}(s_1)S^{-1}(-s_2)(2\pi)^2 \delta^4(s-u) -iK_\up{M}(p,s,u).
\end{gathered}
\end{equation}

As shown in \autoref{Fig-BS-D}, by performing the charge conjugation transformation, the Bethe-Salpeter equation for diquark cores reads
%\begin{equation}\label{E-BS-diquark1}
%\Gamma_\up{D}(p,s)_{ij}=\int \frac{\up{d}^4 u}{(2\pi)^4} iK_\up{D}(s-u)_{im\rq{};jn\rq{}} S_{m\rq{}m}(u_1) \Gamma_\up{D}(p,u)_{mn}  S_{n\rq{}n}(u_2)\,.
%\end{equation}
 %Concequently, we can rewrite \eref{E-BS-diquark1} as
%\begin{equation}\label{E-BS-M1}
%\Gamma^c_{ij}(p,s) = \int \frac{\up{d}^4 u}{(2\pi)^4}  iK^c_{im\rq{};n\rq{}j}(s-u)  \left [ S (u_1) \Gamma^c(p,u)  S (-u_2) \right ]_{m'n'},
%\end{equation}
%where
%\[ iK^c_{im\rq{};n\rq{}j}(s-u) =  \frac{1}{2}i V_\up{M} (s-u) (\gamma^\alpha)_{im\rq{}} (\gamma_\alpha)_{n\rq{}j} =  \frac{1}{2} iK_\up{M} (s-u)_{im\rq{};n\rq{}j}.\]
%Now, in the matrix form, \eref{E-BS-M1} reads
\begin{equation}
\begin{gathered}\label{E-BS-M2}
\Gamma^c(p,s) = i\int \frac{\up{d}^4 u}{(2\pi)^4}  K^c(s-u) [ S (s_1) \Gamma^c(p,s)  S (-s_2)]\,,
\end{gathered}
\end{equation}
where $\Gamma^c \equiv \Gamma_\up{D} {C}^{-1} $, and ${C}\equiv i\gamma^2 \gamma^0$ denotes the charge conjugation operator; $\Gamma_\up{D}$ is the diquark vertex; and $K^c=\frac{1}{2} K_\up{M}$. 
Similarly, the diquark BS wave function can be defined as $\psi^c(p,s) \equiv S (s_1) \Gamma^c(p,s)  S (-s_2)$. Note that \eref{E-BS-M2} and \eref{E-BS-Meson} share exactly the same form with the only difference that the diquark bound strength is halved due to the color factor. Therefore, \eref{E-BS-M2} can be easily solved to obtain mass spectra and wave functions of the diquark cores.

\subsection{\textup{The Salpeter equation and wave function for the $1^+$ diquark core}}
As pointed out in the Introduction, according to the Pauli principle, the heavy diquark core $(bc)$ in color anti-triplet at the ground state ($S$-wave) may have $J^P=1^+$ or $0^+$, while $(cc)$ and $(bb)$ can only be in $J^P=1^+$. In this work, we restrict ourselves to $1^+$ diquark cores $(cc)$, $(bc)$ and $(bb)$. Following the standard procedures in Ref.\,\cite{Salpeter1952}, we define the 3-dimensional Salpeter wave function $\varphi^c(p,s_\perp) \equiv i \int \frac{\up{d}s_p}{2\pi}\psi^c(p,s)$, where $s_p=s\cdot \hat{p}$, $s_\perp=s-s_p \hat{p}$, and $\hat{p}=\frac{p}{\mu}$. Then we can obtain the 3-dimensional Salpeter equation as
\begin{equation}\label{E-BS-HM2}
\mu\varphi^c(p,s_\perp)%=\mathscr{H}\varphi^c(p,s_\perp)
= (e_1+e_2) \hat H(s_{1\perp})\varphi^c(s_\perp) +\frac{1}{2} \left[ \hat H(s_{1\perp}) W(s_\perp) - W(s_\perp) \hat H(s_{2\perp})\right]\,,
\end{equation}
where $\mu$ is the bound state mass; $e_i=\sqrt{\mu_i^2-s_{i\perp}^2}~(i=1,2)$ denotes the quark kinematic energy; $\hat H(s_{i\perp}) \equiv\frac{1}{e_i} (s^\alpha_{i\perp}\gamma_\alpha+\mu_i)\gamma^0$ are the usual Dirac Hamilton divided by $e_i$; $W(s_\perp)\equiv\gamma^0\Gamma^c(p,s_\perp)\gamma_0$ is the potential energy part; and the 3-dimensional vertex is expressed as
\begin{equation}\label{E-vertex-3D}
\Gamma^c(p,s_\perp) =\int \frac{\up{d}^3 u_\perp}{(2\pi)^3}  K^c(s_\perp-u_\perp) \varphi^c(p,u)\,.
\end{equation}
Also we obtain a constraint condition for the Salpeter wave function
\begin{equation} \label{E-BS-constraint}
\hat  H(s_{1\perp}) \varphi^c(s_\perp) + \varphi^c(s_\perp)\hat  H(s_{2\perp}) =0\,,
\end{equation}

Accordingly, the normalization condition now becomes
\begin{equation}
%\int \frac{\up d^3 \vec q}{(2\pi)^3}\frac{1}{2} \up{Tr}\left[ \varphi^\dagger  \hat H_1\varphi -  \varphi \hat H_2\varphi^\dagger \right]=
\int \frac{\up d^3 \vec s}{(2\pi)^3}  \up{Tr}~ \varphi^{c\dagger}(p,s_\perp)  \hat H(s_{1\perp})\varphi^c(p,s_\perp)=2\mu.
\end{equation}

Since a diquark consists of two quarks, its parity is just opposite to its meson partner. Then, the wave function of the diquark with $J^P=1^+$ can be decomposed as
\begin{equation}\label{E-1+wave}
\varphi^c(1^+)= e\cdot \hat s_\perp \left(f_1+f_2 \frac{\sl p}{\mu}+f_3 \frac{\sl s_\perp}{s}+f_4 \frac{\sl p\sl s_\perp}{\mu s}\right) +  i\frac{ \epsilon_{\alpha p s_{\!\perp} e}}{s\mu} \gamma^\alpha \left(f_5  \frac{\sl p \sl s_\perp}{\mu s}+  f_6 \frac{\sl s_\perp}{s} + f_7  \frac{\sl p}{\mu}+ f_8  \right) \gamma^5,
\end{equation}
where $s=\sqrt{-s_\perp^2}$ and $\hat s_\perp=\frac{s_\perp}{s}$; $e$ is the polarization vector fulfilling the Lorentz condition $e^\alpha p_\alpha=0$. Note that there are eight radial wave functions $f_i~(i=1,2,\cdots,8)$, but only four of them are independent due to the constraint condition \eref{E-BS-constraint}, i.e.,
\begin{equation}
f_1=-A_{+} f_3,~~f_4=-A_-f_4,~~f_7=A_- f_5,~ ~ f_8=A_+ f_6,
\end{equation}
where $A_\pm  \equiv \frac{s(e_1\pm e_2)}{\mu_1e_2+ \mu_2e_1}$.
%\begin{equation*}
%\beta_+  \equiv \frac{s(\epsilon_1+\epsilon_2)}{\mu_1\epsilon_2+ \mu_2\epsilon_1}, ~~\beta_- \equiv \frac{s(\epsilon_1-\epsilon_2)}{ \mu_1\epsilon_2+\mu_2\epsilon_1}.
%\end{equation*}
Note that the meson wave functions with $J^{PC}=1^{--}$ share the same form of \eref{E-1+wave}.
Inserting the decomposition into \eref{E-BS-HM2} and taking the Dirac trace, we can obtain four coupled eigenvalue equations, which are explicitly shown in appendix\ref{App-1}.
The normalization can then be simply expressed as,
\begin{equation}
\int \frac{\up{d}^3 \vec s}{(2\pi)^3} \frac{8e_1e_2}{3\mu(\mu_1e_2+\mu_2e_1)}\left[f_3(s)f_4(s)-2f_5(s)f_6(s)\right]=1.
\end{equation}

\begin{figure}[h!]
\vspace{0.5em}
\centering
\subfigure[Radial wave function for the $(cc)$ ground state.]{\includegraphics[width=0.48\textwidth]{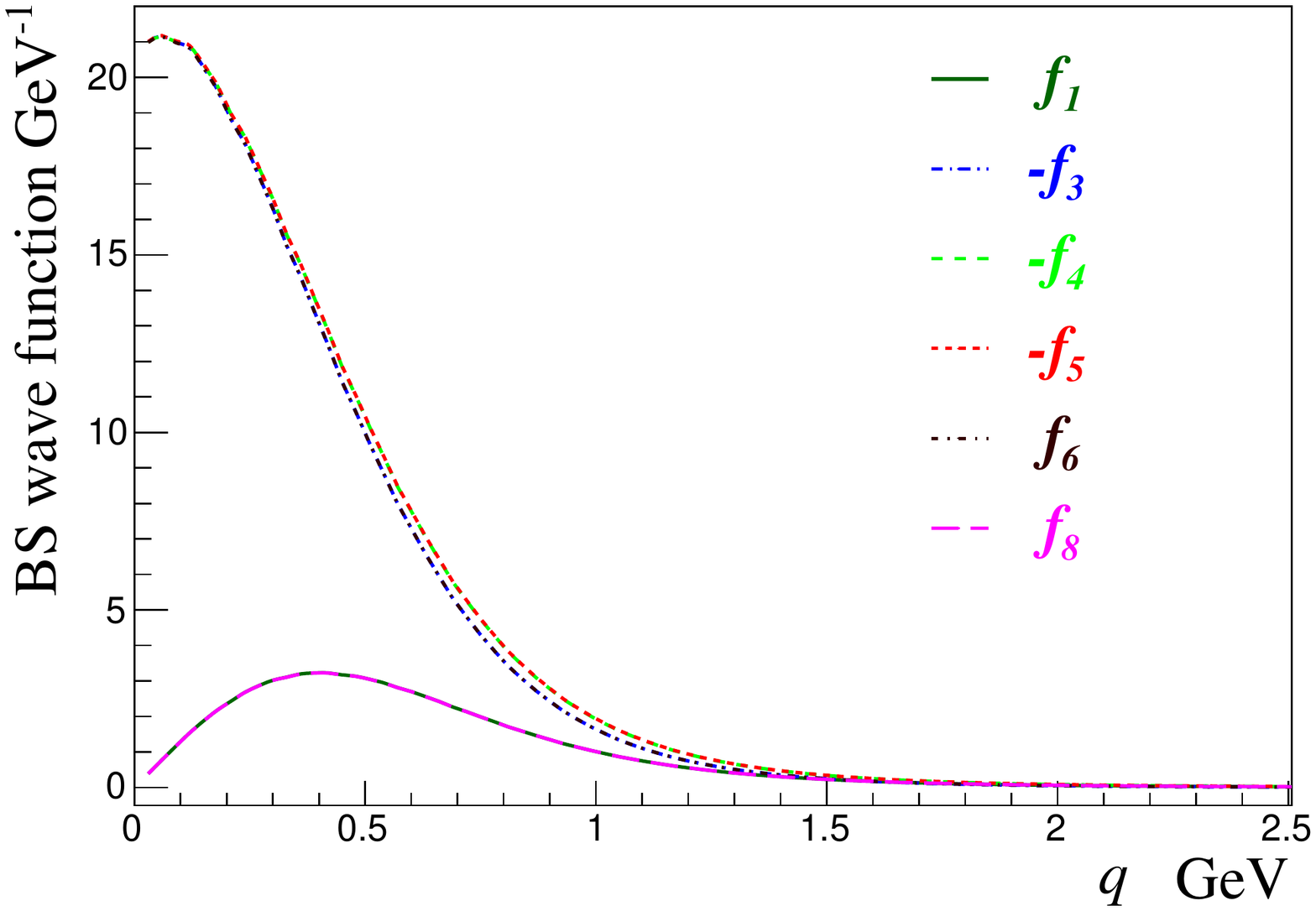} \label{Fig-ccD-n1}}
\subfigure[Radial wave function for the $(cc)$ 1st excited state.]{\includegraphics[width=0.48\textwidth]{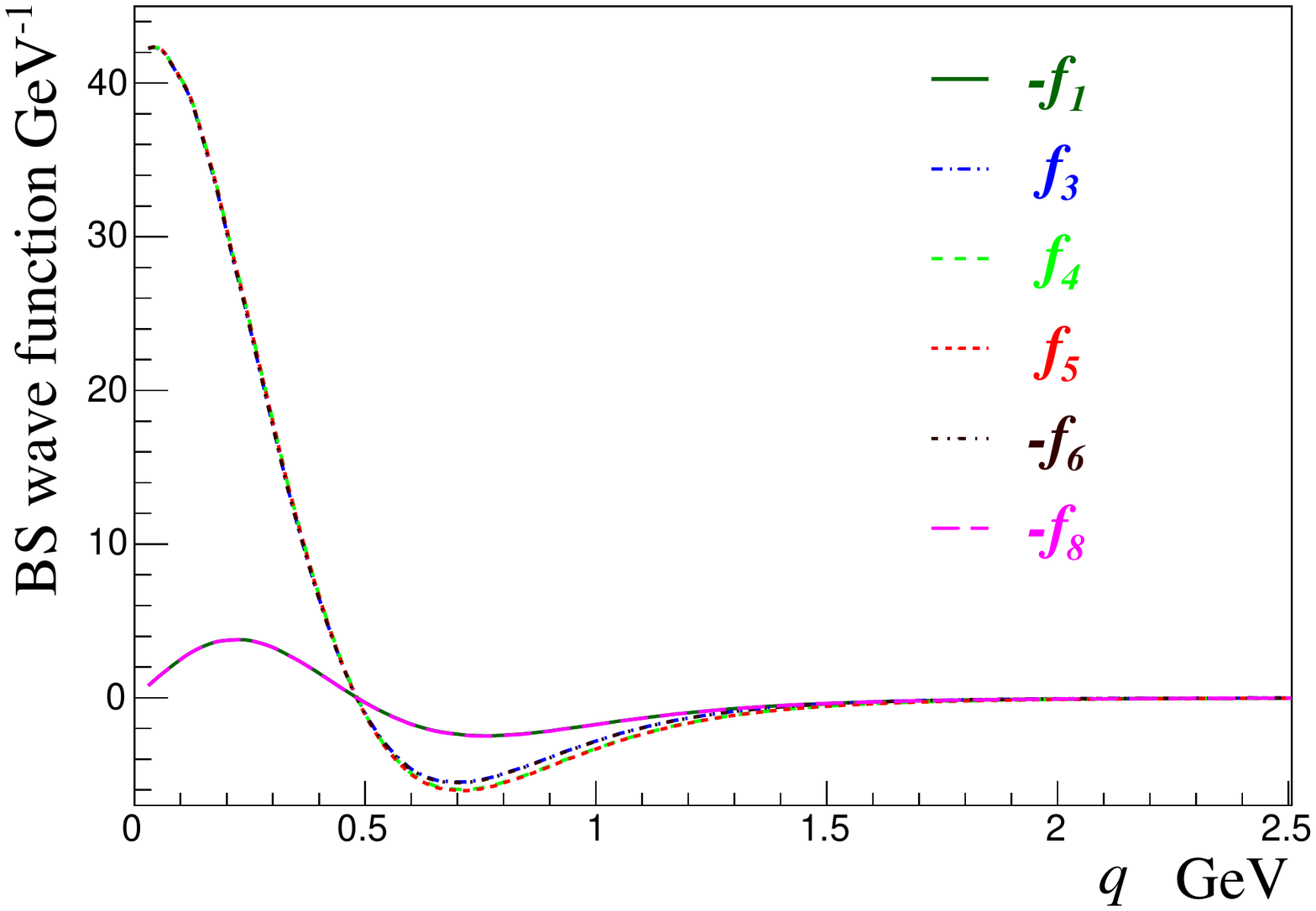} \label{Fig-ccD-n2}}\\
\caption{Radial wave functions for the $(cc)$-diquark core with $J^P=1^+$. }\label{Fig-wave-1+D}
\vspace{0.5em}
\end{figure}

Solving the coupled equations numerically, the mass spectra and wave functions of the diquark cores are obtained.
%In this work, the parameters in the potential are fixed by fitting the mass spectra of the $1^{--}$ $c\bar c$.
The parameters
\[
a =e=2.7183,~~~ \lambda =0.21~\si{GeV}^2, ~~~ \Lambda_\text{QCD} =0.27~\si{GeV},  ~~~a_1 =a_2=0.06~\si{GeV},
\]
and the constituent quark masses
\[
m_u =0.305~\si{GeV},~~~ m_d =0.311~\si{GeV}, ~~~ m_s =0.5~\si{GeV}, ~~~ m_c =1.62~\si{GeV}, ~~~m_b=4.96~\si{GeV}
\]
used in this work are just the same as those adopted in the previous studies and determined by fitting to the heavy and doubly heavy mesons spectra
\,\cite{Chang2010,WangZ2012A,WangT2013,WangT2013A,LiQ2016,LiQ2017,LiQ2017A}. The free parameter $V_0$ in the diquark system is fixed by fitting the mass of its meson partner to the experimental data, e.g., $V_0$ of the $1^+$ $(cc)$-diquark is determined by the $J/\psi$. In this work, we obtain $V_0=-0.221~\si{GeV}$ for $1^+$ $(cc)$, $V_0=-0.147$ GeV for $1^+$ $(bc)$, and $V_0=-0.026$ GeV for $1^+$ $(bb)$.
\begin{table}[h!]
\caption{Mass spectra of $J^P=1^+$ color anti-triplet diquark cores $(cc)$, $(bc)$ and $(bb)$ in units of GeV.}\label{Tab-Mass-cc}
\vspace{0.2em}\centering
\begin{tabular}{ cccccccccc }
\toprule[2pt]
%$c\bar c$ 	&$J/\psi$			&$\psi(2S)$		&$\psi(3770)$		&$\psi(4040)$	&$\psi(4160)$	&$\psi(4260)$	&$\psi(4360)$	 &$\psi(4415)$	&$\psi(4660)$	\\
%\midrule[1.5pt]
%Exp. 		&$3.097$			&$3.686$			&$3.773$			&$4.039$		&$4.191$		&$4.251$		&$4.361$	 	 &$4.421$		&$4.664$		\\
%This 	  	&$3.097$			&$3.686$			&$3.774$			&$4.074$		&$4.128$		&$-$			&$4.374$	 	 &$4.411$		&$4.619$		\\
%\midrule[1.5pt]
%$c c$ 	&$1S$				&$2S$				&$1D$				&$3S$			&$2D$			&$4S$			&$3D$	 		 &$5S$	&$4D$	\\
%\midrule[1.5pt]
%This 		&$3.298$			&$3.631$			&$3.681$			&$3.850$		&$3.881$		&$4.016$		&$4.037$	 	 &$4.150$		&$4.163$		\\
$n_\up{D}L_\up{D}$	&$1S$				&$2S$				&$1D$				&$3S$\\%			&$2D$			&$4S$			&$3D$			 &$5S$\\
\midrule[1.5pt]
$(cc)$ 		&$3.303$			&$3.651$			&$3.702$			&$3.882$\\%		&$3.914$		&$4.059$		&$4.081$	 	 &$4.202$\\
$(bc)$ 		&$6.594$			&$6.924$			&$6.980$			&$7.142$\\%		&$7.178$		&$7.311$		&$7.335$	 	 &$7.448$\\
$(bb)$ 		&$9.830$			&$10.154$		&$10.217$		&$10.361$\\%	&$10.401$	&$10.520$	&$10.548$	 &$10.651$\\
\bottomrule[2pt]
\end{tabular}
\end{table}
The obtained mass spectra are listed in \autoref{Tab-Mass-cc}; and the radial wave functions of $J^P=1^+$ $(cc)$-diquark core, as an example, are shown in \autoref{Fig-wave-1+D}, where $f_2=f_7=0$ are omitted.

The wave functions of the diquark cores at origin are very useful in many applications and defined as
\begin{equation}
\psi(\vec x)\mid_{\vec x=\vec 0}=\int \frac{\up{d}^3 \vec s}{(2\pi)^3}  \varphi^c(p,s_\perp) \equiv \psi_\up{V}\sl e+\psi_\up{T} \sl e \hat{\sl{p}},
\end{equation}
where $\psi_\up{V}$ and $\psi_\up{T}$ are related to the Salpeter radial wave functions by
\begin{equation}\notag
\begin{aligned}
\psi_\up{V} &
=-\frac{1}{4}\int \frac{\up{d}^3 \vec s}{(2\pi)^3} \up{Tr}~ \varphi^c(|\vec s\,|) \sl e
  = -\frac{1}{3}\int \frac{\up{d}^3 \vec s}{(2\pi)^3}(f_3+2f_5),\\
\psi_\up{T} &
=-\frac{1}{4 }\int \frac{\up{d}^3 \vec s}{(2\pi)^3}\up{Tr}~ \varphi^c(|\vec s\,|) \hat{\sl p}\sl e
= +\frac{1}{3}\int \frac{\up{d}^3 \vec s}{(2\pi)^3}(f_4-2f_6)\,.
\end{aligned}
\end{equation}
The obtained numerical values of $\psi_\up{V}$ and $\psi_\up{T}$ are listed in \autoref{Tab-zero}.
\begin{table}[h!]
\caption{Wave functions at origin of $1^+$ diquark cores $(cc)$, $(bc)$ and $(bb)$ for the ground states (in GeV$^2$).}\label{Tab-zero}
\vspace{0.2em}\centering
\begin{tabular}{ cccccccccc }
\toprule[2pt]
diquark cores	&$(cc)$				&$(bc)$				&$(bb)$\\
\midrule[1.5pt]
$\psi_\up{V}$ 		&$0.155$			&$0.299$			&$0.618$\\	
$\psi_\up{T}$ 		&$-0.144$			&$-0.287$			&$-0.600$\\
\bottomrule[2pt]
\end{tabular}
\end{table}

\subsection{\textup{The form factor of the diquark core coupling to a gluon}}

\begin{figure}[h!]
\centering
\subfigure[A gluon couples with quark-2 in the diquark.]   {\includegraphics[width=0.45\textwidth]{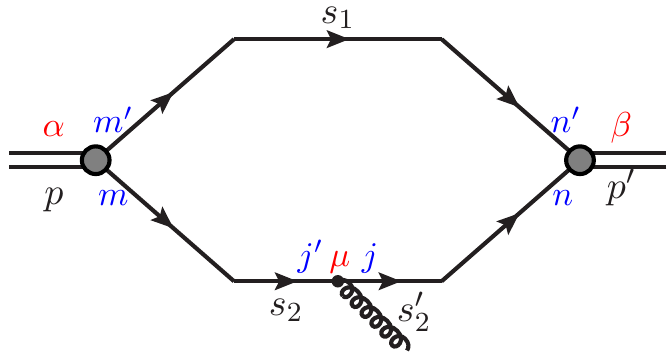} \label{Fig-diquark-form1}}
\subfigure[A gluon couples with quark-1 in the diquark.]   {\includegraphics[width=0.45\textwidth]{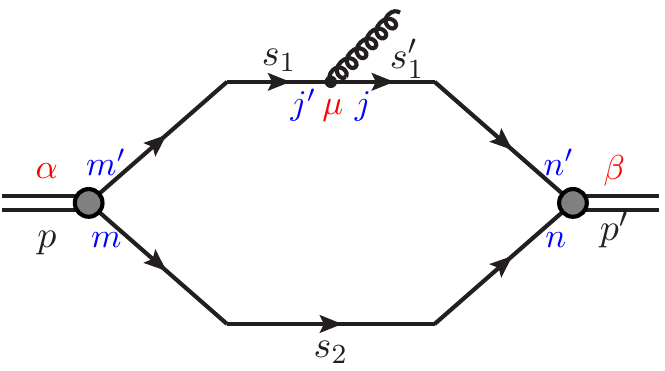} \label{Fig-diquark-form2}}
\caption{The vertex diagram of a diquark core to a gluon.} \label{Fig-diquark-form}
\end{figure}

In the BSEs of mesons or diquarks, the constituent quarks (antiquarks) are the point-like particles. But in the BSEs of baryons in terms of quark and diquark, the structures of the non-pointlike diquarks have to be considered. The coupling between diquarks and gluons is modified by form factors. Thus, in this subsection, we will discuss the relevant form factor within the Mandelstam formulation.

The Feynman diagrams of a doubly heavy diquark core coupling to a gluon are shown in \autoref{Fig-diquark-form}. Since the coupling vertex $\Sigma^{\alpha\beta\mu}$, as a `vector current matrix element', is conserved, it can be decomposed by three independent form factors $\sigma_i\, (i=1,3,5)$ as
\[
\Sigma^{\alpha\beta\mu}=-\sigma_1(t^2) g^{\alpha\beta} (p+p')^\mu + \sigma_{3}(t^2)  (p^\beta g^{\alpha\mu} + p'^\alpha g^{\beta\mu})  + \sigma_{5}(t^2) p'^\alpha p^\beta (p^\mu + p'^\mu),% \\
%&\overset{\up{IA}}{\simeq}  \sigma_1(t) g^{\alpha\beta} (p_1 + k_1)^\mu,
\]
where $\sigma_i$ only depend on the transfer momentum $t^2 \equiv (p-p')^2$. Note that, in the instantaneous approximation, $p^0=p'^0$ and hence $t^2=(p_\perp-p'_\perp)^2$ is always space-like. Considering the fact that the contributions of $\sigma_3$ and $\sigma_5$ to the BSEs of baryons is small compared with $\sigma_1$, we will keep the dominant $\sigma_1$ only and omit the subscript for simplicity.

Corresponding to the Feynman diagrams in \autoref{Fig-diquark-form}, the form factor consists of two terms
\begin{equation}\label{eqFF}
\begin{aligned}
\Sigma^{\alpha\beta\mu}&=\frac{1}{2}\left(\Sigma^{\alpha\beta\mu}_1+\Sigma^{\alpha\beta\mu}_2 \right),
\end{aligned}
\end{equation}
where the factor $\frac{1}{2}$ is due to the normalization condition and guarantees the form factor $\sigma(t^2)=1$ at zero transfer momentum
($t^2=0$). Explicitly, for example, the amplitude $\Sigma^{\alpha\beta\mu}_1$ corresponding to \autoref{Fig-diquark-form1} reads
\begin{equation}\notag
\begin{aligned}
\Sigma^{\alpha\beta\mu}_1%&=-\int \frac{\up{d}^4 s}{(2\pi)^4} \up{Tr}~\bar{\Gamma}^\beta_\up{D}(p',s')S(s_1)\Gamma_\up{D}^\alpha(p,s) S^\up{T}(s_2) (\gamma^\mu)^\up{T} S^\up{T}(s'_2)\\
&=-\int \frac{\up{d}^4 s}{(2\pi)^4} \up{Tr}~\bar{\Gamma}^\beta_c(p',s') S(s_1)\Gamma^\alpha_c(p,s) S(-s_2) \gamma^\mu S(-s'_2)\\
&\simeq \int \frac{\up{d}^3 \vec s}{(2\pi)^3} \up{Tr}~\bar{\varphi}^\beta_c(p',s'_{\!\perp})\gamma^0 \varphi^\alpha_c(k_1,s_\perp) \gamma^\mu\,,
\end{aligned}
\end{equation}
where the contour integration over $s^0$ is performed and only the dominant contribution is kept. And the amplitude of \autoref{Fig-diquark-form2} can be easily obtained by the relation $\Sigma^{\alpha\beta\mu}_2= \Sigma^{\alpha\beta\mu}_1$ with $(\mu_1\rightleftharpoons \mu_2)$.

\begin{figure}[h!]
\centering
\subfigure[Form factor of the ground diquark.]   {\includegraphics[width=0.46\textwidth]{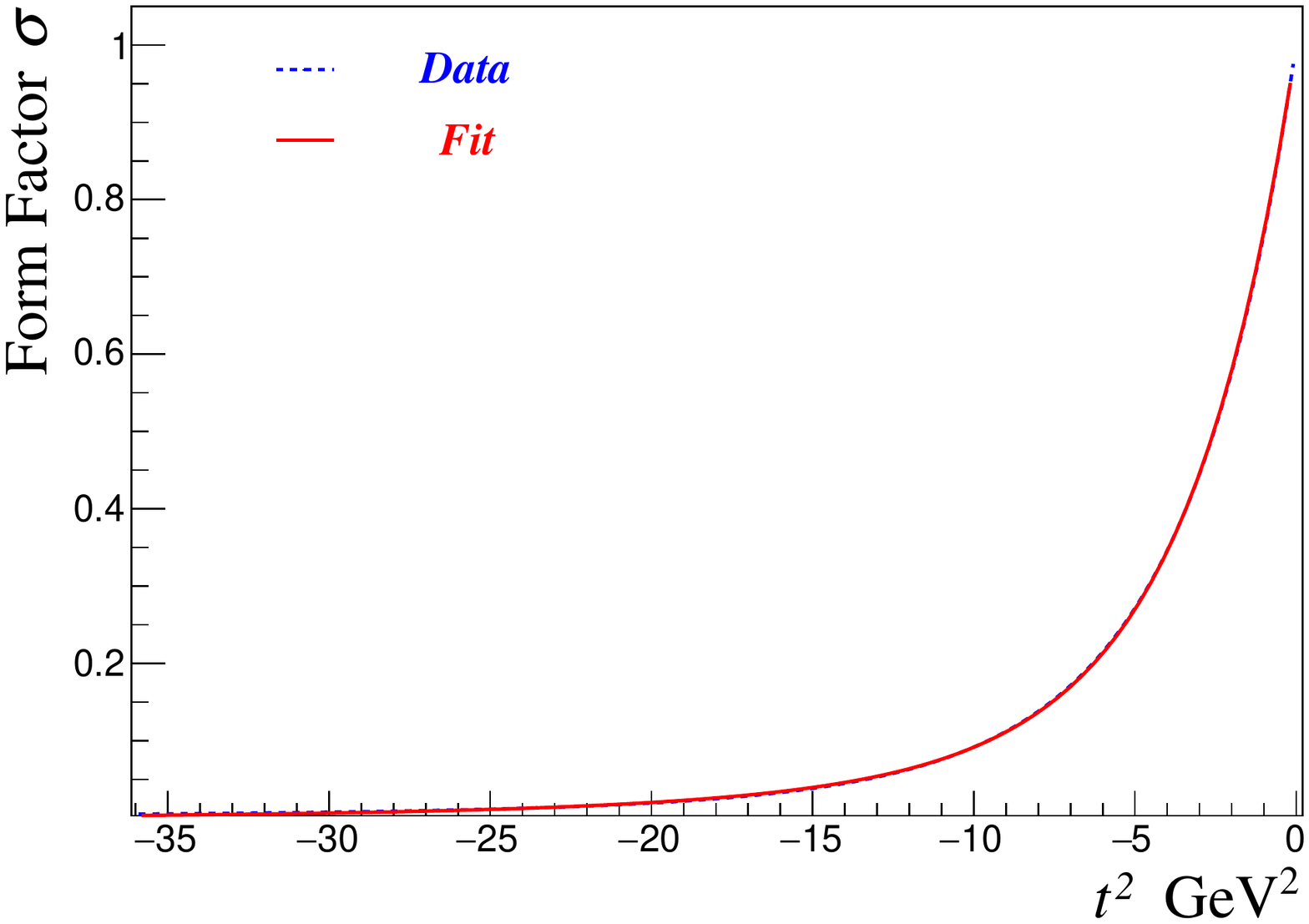} \label{Fig-form-n1}}
\subfigure[Form factor of the first excited diquark.]   {\includegraphics[width=0.46\textwidth]{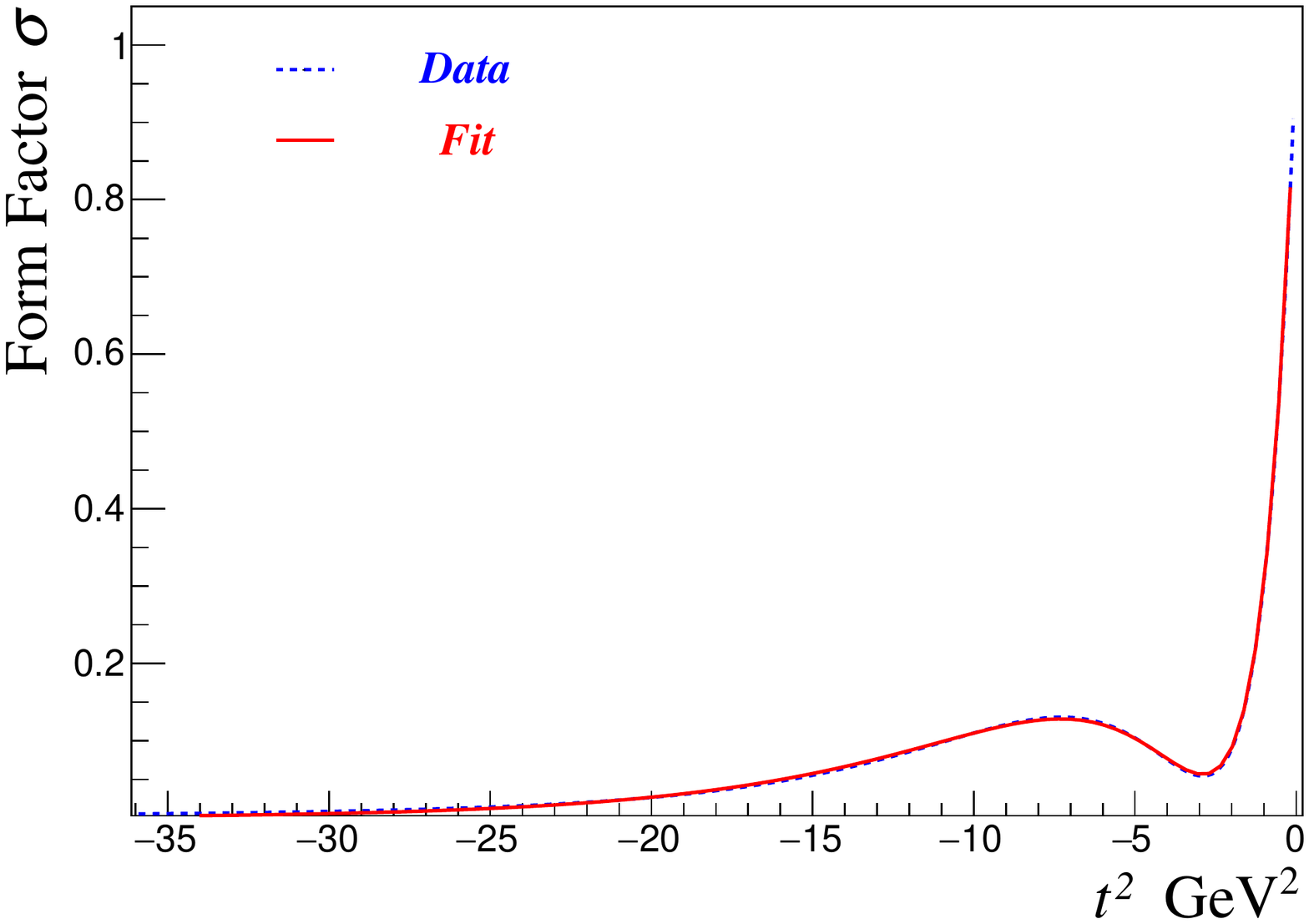} \label{Fig-form-n2}}
\caption{Form factor of the $1^+$ $(cc)$-diquark core coupling to a gluon. }\label{Fig-Xicc-form}
\end{figure}

Inserting the $1^+$ Salpeter wave function \eref{E-1+wave} into \eref{eqFF}, we can obtain $\sigma$ for $1^+$ $(cc)$-diquark in the ground and first excited states, which are shown in \autoref{Fig-Xicc-form}.  For convenience, we can parameterize the obtained numerical form factor $\sigma$ as
\begin{equation}
\sigma(t^2)=Ae^{\kappa_1t^2}+(1-A)e^{\kappa_2t^2}\,.
\end{equation}
For example, we obtain $A=0.162, ~\kappa_1=0.109, ~\kappa_2=0.312$ for the ground state. 

\section{Doubly heavy baryon as the bound state of the diquark core and a light quark}\label{Sec-3}

In this section, we first construct the Bethe-Salpeter equation of a $1^+$ diquark core and a light quark, and then derive the 3-dimensional (Bethe-)Salpeter equation in the instantaneous approximation. At last, we compute mass spectra and wave functions for baryons with the quantum numbers $J^P=\frac{1}{2}^+$ and $\frac{3}{2}^+$.

\begin{figure}[h!]
\centering
\includegraphics[width = 0.9\textwidth, angle=0]{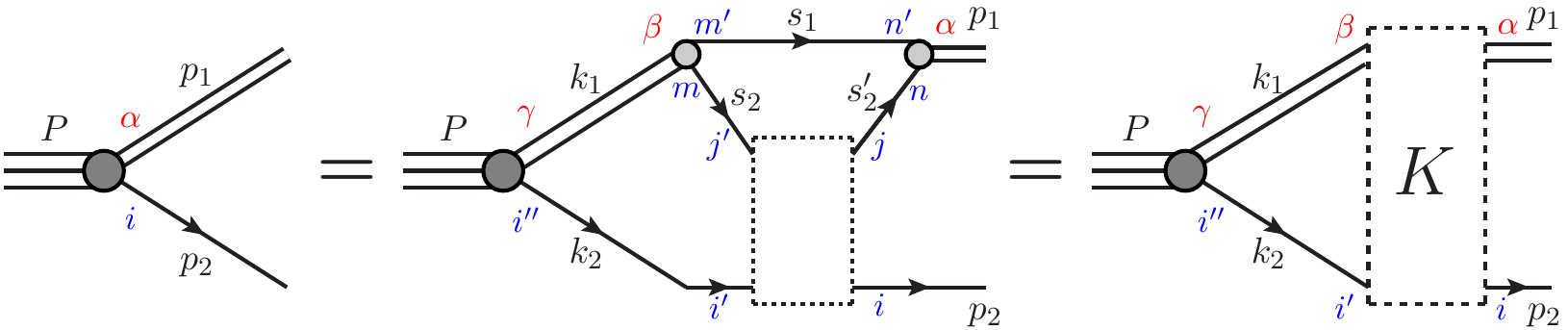}
\caption{Bethe-Salpeter equation of the baryon based on the diquark model.  The Greeks~(red) are used for the Lorentz indices; the Romans (blue), the Dirac indices. $P,~p_1(k_1),~p_2(k_2)$ denote the momenta of the baryon, heavy diquark and the third light quark respectively.}\label{Fig-BS-B}
\end{figure}

\subsection{\textup{The Bethe-Salpeter equation of a baryon with the $1^+$ heavy diquark core}}

The BS equation for a baryon with the $1^+$ heavy diquark core is schematically depicted by \autoref{Fig-BS-B}, which can be expressed using the matrix notation as
\begin{equation}\label{E-BS-BD1}
\Gamma^\alpha(P,q,r)=\int  \frac{\up{d}^4 k}{(2\pi)^4} (-i)K^{\alpha\beta}(p_1,k_1;p_2,k_2)  \left[S(k_2)\Gamma^{\gamma} (P,k,r) D_{\beta\gamma}(k_1) \right],
\end{equation}
where $\Gamma^\alpha(P,q,r)$ is the BS vertex of the baryon; $P$, the baryon momentum with $P^2=M^2$ and $M$ denoting the baryon mass; $r$, the baryon spin state; $(-i)K(p_1,k_1;p_2,k_2)$ represents the effective diquark-quark interaction kernel; and $D_{\beta\gamma}(k_1)=i\frac{-g^{\beta\gamma}+ p^\beta_1p^\gamma_1/m_1^2}{p_1^2-m_1^2+i\epsilon}\,$ is the free diquark propagator (axial-vector particle) with mass $m_1$, and here we do not consider the self-energy correction of the doubly heavy diquark; the internal momenta $q$ and $k$ are defined as
\[
q=\alpha_2 p_1-\alpha_2 p_2,~~~~k=\alpha_2 k_1-\alpha_1 k_2,
\]
with $\alpha_i=\frac{m_i}{m_1+m_2}$ and $m_2$ representing the constituent mass of the third quark.  From now on, the symbols $P$ and $r$ in the BS vertex $\Gamma^\alpha(P,q,r)$ would be omitted unless it is necessary. 

In terms of the diquark form factor, the baryon effective interaction kernel can be simply expressed as
\begin{equation}\label{E-Kb3}
(-i)K^{\alpha\beta}(p_1,k_1;p_2,k_2)=- iV_\up{M}(k-q)\Sigma^{\alpha\beta\mu} \gamma_\mu \,.
\end{equation}
In principle, the form factors must be calculated for off-shell diquarks. In this work, we simply generalize the on-shell form factors obtained in the previous section to the off-shell ones, namely, $\Sigma^{\alpha\beta\mu}=\sigma(t^2) g^{\alpha\beta}h^\mu$ and $h=k_1+p_1$. Notice that the diquark-quark potential is now modified by the form factor, which reflects the structure effect of the nonpoint-like diquark.

The BS wave function $B_\alpha(P,q)$ of the baryon can be defined as
\begin{equation}
B_\alpha(P,q)=S(p_2) D_{\alpha\beta}(p_1) \Gamma^{\beta}(P,q),
\end{equation}
with the constraint condition $P^\alpha B_{\alpha}(P,q)=0$. 
%The BS vertex can be expressed by the BS wave function as,
%\begin{align}\label{E-BS-BD1}
%\Gamma^\alpha(P,q)=\int  \frac{\up{d}^4 k}{(2\pi)^4}(-)iK^{\alpha\beta}(p_1,k_1;p_2,k_2) B_\beta (P,k).
%\end{align}
%Then the baryon Bethe-Salpeter \eref{E-BS-BD1} can be further expressed as an integral equation of the wave function,
%\begin{equation}
%B_\alpha(P,q) = D_{\alpha\beta}(p_1) S(p_2)\int \frac{\up d^4 k}{(2\pi)^4}(- )i K^{\beta\gamma} B_\gamma(P,k) .
%\end{equation}
%This is the fundamental formula we obtained for the doubly heavy baryon with $1^+$ diquark core.
Accordingly, the normalization condition can be expressed as,
\begin{equation} \label{E-Norm-B4D}
-i\int \int \frac{\up{d}^4 q}{(2\pi)^4}\frac{\up{d}^4 k}{(2\pi)^4} ~  \bar{B}_\alpha(q,\bar r) \frac{\partial}{\partial P^0} \left[ I^{\alpha\beta}(P,k,q) \right] B_\beta(k,r)  =2M \delta_{r\bar r},
\end{equation}
where the operator $I^{\alpha\beta}(P,k,q)$ has the following form,
\begin{equation}
I^{\alpha\beta}(P,k,q)=S^{-1}(p_2)D_{\alpha\beta}^{-1}(p_1)(2\pi)^2 \delta^4(k-q) + iK^{\alpha\beta}(p_1,k_1;p_2,k_2).
\end{equation}

\subsection{\textup{BSE of the doubly heavy baryons in instantaneous approximation}}
In the instantaneous approximation, the factor $h^0=2(\alpha_1M+q_P)$ depends on $q_P$ and $M$ explicitly, which is different from the case of mesons or diquarks. %For doubly heavy baryons, the spatial components of the kernel are suppressed by $v/c$. 
As we will see, the factor $h^0$ plays an important role in the derivation of the three-dimensional Salpeter equation of baryons. %By factoring out the Lorentz indices, the kernel in \eref{E-Kb3} can be expressed as
%\begin{equation} \notag
%\begin{aligned}
%(-i)K^{\alpha\beta}(p_1,k_1)&= -i\sigma(t^2) V_\up{M}(k_\perp-q_\perp) g^{\alpha\beta} h^0  \gamma_0\\
%	&\equiv(-i) K(p_1,k_1) g^{\alpha\beta}.
%\end{aligned}
%\end{equation}
%The baryon vertex $\Gamma^\alpha(P,q)$ now satisfies the following equation,
%\begin{equation} \label{E-Vertex-B4}
%\Gamma^\alpha (P,q)=-i\int \frac{\up d^4 k}{(2\pi)^4} K(p_1,k_1) B^\alpha(P,k).
%\end{equation}
First, we introduce the instantaneous baryon kernel as
\begin{equation} \notag
K(k_\perp-q_\perp)=\sigma(t^2) V_\up{M}(k_\perp-q_\perp) \gamma_0\,,
\end{equation}
and then have $K^{\alpha\beta}(p_1,k_1) = h^0 g^{\alpha\beta} K(q_\perp-k_\perp) $.
Then we define the baryon Salpeter wave function as
\begin{equation}\notag
\varphi_\alpha(q_\perp) \equiv -i\int \frac{\up d q_P}{2\pi} B_\alpha(q),
\end{equation}
with the constraint condition $P^\alpha \varphi_\alpha =0$.
%Now we have
%\begin{equation}
%	\Gamma_\alpha(P,q)=h_0 \Gamma_{\alpha}(q_\perp)\,,
%\end{equation}

Now the baryon BS wave function can be expressed as
\begin{equation}\label{E-BS-BD1-1}
B^\alpha(q)=h^0 S(p_2)  D^{\alpha\beta}(p_1) \Gamma_\beta (q_\perp)\,,
\end{equation}
where the three-dimensional vertex $\Gamma_a(q_\perp)$ is expressed by the Salpeter wave function as
\begin{equation}
\Gamma_{\alpha} (q_\perp) \equiv \int \frac{\up d^3 k_\perp}{(2\pi)^3}K(k_\perp-q_\perp) \varphi_\alpha(k_\perp).
\end{equation}
Due to the constraint condition $P_\alpha \varphi^\alpha(P)=0$, the components of $D^{\alpha\beta}(p_1)$ parallel to $P$ vanish. Hence we can rewrite $D^{\alpha\beta}(p_1)$ as
\begin{equation}\notag
D^{\alpha\beta}(p_1)=i\frac{\vartheta^{\alpha\beta}}{p_1^2-m_1^2+i\epsilon},~~~\vartheta^{\alpha\beta}\equiv -g^{\alpha\beta}+\frac{p_{1\perp}^\alpha  p_{1\perp}^\beta}{m_1^2}.
\end{equation}
Note that in Ref.\,\cite{Keiner1996}, the term $i\frac{p^\alpha_1 p^\beta_1/m_1^2}{p_1^2-m_1^2}$ is simply neglected.

We follow Salpeter's method in Ref.\,\cite{Salpeter1952}. Performing the contour integral over $q_P$ on both sides of \eref{E-BS-BD1-1} (see appendix\ref{App-SE} for details), we eventually obtain the Schr\"{o}dinger-type Salpeter equation for baryons with $1^+$ diquark cores as
\begin{equation} \label{E-BSB-D1-2}
M\varphi^\alpha(q_\perp)=(\w_1+\w_2)\hat H( p_{2\perp})\varphi^\alpha(q_\perp) + \hat{H}(p_{2\perp}) \gamma^0 \vartheta^{\alpha\beta}\Gamma_\beta (q_\perp) \,,
\end{equation}
where the baryon mass $M$ behaves as the eigenvalue; the first term expresses the contribution to the mass from the kinetic energy, and the second term that from the potential energy. Correspondingly, the normalization condition can be expressed as (see appendix\ref{App-2} for details)
\begin{equation}\label{E-Norm-D1}
\int \frac{\up{d}^3 q_\perp}{(2\pi)^3} \bar\varphi_\alpha(P,q_\perp,\bar r) \gamma^0 2\left( \alpha_1 M\hat H( p_{2\perp})+ \w_q \right) d^{\alpha\beta} \varphi_\beta(P,q_\perp,r)=2M\delta_{r\bar r},
\end{equation}
with $\w_q\equiv \alpha_2\w_1-\alpha_1\w_2$ and $d^{\alpha\beta}\equiv-g^{\alpha\beta}-\frac{p^\alpha_{1\perp} p^\alpha_{1\perp}}{\w_1^2}$.

\subsection{\textup{The Salpeter wave functions for baryons with $J^P=\frac{1}{2}^+$ and $\frac{3}{2}^+$}}
In this subsection, we construct the baryon wave functions directly from the good quantum number $J^P$. Then the $D$-wave components are automatically included and the possible $S$-$D$ mixing can also be obtained directly by solving the corresponding BSE. 

For ground states with $L=0$, where $L$ denotes the diquark-quark orbital angular momentum, the $1^+$ diquark and the $\frac{1}{2}^+$ quark can form a baryon doublet $(\frac{1}{2},\frac{3}{2})^+$. For the wave function of $J^P=\frac{1}{2}^+$ baryons, analyzing its total angular momentum, Lorentz and Dirac structures, we can decompose it as
\begin{equation}\label{E-wave-d1-1H}
\begin{aligned}
\varphi_\alpha(P,q_\perp,r)&=\left(g_1 + g_2\frac{\sl q_\perp}{q}\right)\xi_{1\alpha} \gamma^5 u(P,r)+\left(g_3+g_4 \frac{\sl q_\perp}{q}\right)\hat{q}_{\perp\alpha}\gamma^5 u(P,r)\\
&=A_{\alpha} u(P,r),
\end{aligned}
\end{equation}
where we have defined $A_\alpha$ to factor out the spinor; $\xi_{1\alpha}=(\gamma_\alpha+\frac{P_\alpha}{M})$, and $q=\sqrt{-q_\perp^2}$; the radial wave functions $g_i~(i=1,2,3,4)$ explicitly depend on $q$; and $u(P,r)$ is the Dirac spinor with momentum $P$ and spin state $r$. The conjugation is defined as usual $\bar \varphi_\alpha(P,q_\perp,r)=\varphi_\alpha^\dagger \gamma_0$. 
%\begin{equation}\notag
%A_\alpha \equiv \left(g_1 + g_2\frac{\sl q_\perp}{q}\right)\xi_{1\alpha} +\left(g_3+g_4 \frac{\sl q_\perp}{q}\right)\xi_{2\alpha} \,.
%\end{equation}
%Then the spinor in \eref{E-wave-d1-1H} can be factored out
%\begin{equation}\label{E-wave-1-2A}
%\varphi_\alpha(q_\perp,r)=A_\alpha u(P,r).
%\end{equation}
Note that in Ref.~\cite{WengMH2011} only $g_1$ and $g_2$ are included. In fact, the full Salpeter wave function should have four independent terms, the `small components' $g_3$ and $g_4$ correspond to part of the $P$- and $D$-wave components respectively, and play certain roles for the exact solutions of the $J^P=\frac{1}{2}^+$ baryons, especially, they will become important for the excited states. To see the partial-wave components more clear,  we rewritten the Salpeter wave function of $\frac{1}{2}^+$ in terms of the spherical harmonics as, 
\begin{equation} \label{E-1-2-Ylm}
\begin{aligned}
\varphi_\alpha &=C_0  Y_0^0 \left(g_1 \xi_{1\alpha} - g_4\frac{ \gamma_\alpha}{3} \right)\gamma^5u  + C_1\left(Y_1^0\Gamma_{\alpha3}-Y_1^{-1} \Gamma_{\alpha+} -  Y_1^{1}\Gamma_{\alpha-}  \right) \gamma^5 u  \\
&+C_2g_4\left[Y_2^{-2}g_{\alpha+} \gamma_+ -Y_2^{-1}\frac{ (g_{\alpha3}\gamma_+  + g_{\alpha+}\gamma_3 )}{\sqrt{2}}  +Y_2^0 \frac{(g_{\alpha+}\gamma_-+g_{\alpha-}\gamma_++2 g_{\alpha3}\gamma_3)}{\sqrt{6}} \right. \\
& \left. - Y_2^1 \frac{(g_{\alpha3}\gamma_-+g_{\alpha-} \gamma_3 )}{\sqrt{2}} + Y_2^2 g_{\alpha-} \gamma_- \right] \gamma^5 u,
\end{aligned}
\end{equation}
where the coefficients are $C_0=2\sqrt{ \pi} $, $C_1=\frac{1}{\sqrt{3}}C_0$, and $C_2=\sqrt{\frac{2}{15}}C_0$;  $\gamma_{\pm} = \mp\frac{1}{\sqrt{2}} (\gamma_1\pm i \gamma_2)$; $\Gamma_{\alpha{\pm}} = \mp \frac{1}{\sqrt{2}}(\Gamma_{\alpha1}\pm i \Gamma_{\alpha2})$, and $\Gamma_{\alpha i} \equiv (g_2 \gamma_i \xi_{1\alpha} + g_3 g_{\alpha i})$ with $i=1,2,3$; $g_{\alpha\pm} = \mp\frac{1}{\sqrt{2}} (g_{\alpha1}\pm i g_{\alpha2})$ with $g_{\alpha\beta}$ denoting the Minkowski metric tensor; $Y_l^m$ is the usual spherical harmonics.

From the above decomposition \eref{E-1-2-Ylm}, we can conclude that the Salpeter wave function for  $\frac{1}{2}^+$ baryon contains the $S$-, $P$-, and $D$-wave components, namely,: the $g_1$ part corresponds to the $S$-wave; $g_{2}$ and $g_3$ parts corresponds to the $P$-wave; and $g_4$ part makes contribution to both the $S$- and $D$-wave components. The numerical details will be presented in the next section.

With the specific wave function in \eref{E-wave-d1-1H}, the Salpeter \eref{E-BSB-D1-2} can be expressed as
\begin{equation} \label{E-eigen-D1-1H}
MA_{\alpha}u(P,r)=(\w_1+\w_2)\hat H( p_{2\perp})A_{\alpha} u(P,r) + \vartheta_{\alpha\beta}\hat{H}(p_{2\perp}) \gamma^0 \int \frac{\up d^3 k_\perp}{(2\pi)^3} K(k_\perp-q_\perp) A^{\beta}u(P,r).
\end{equation}
Projecting the radial wave functions out, we arrive at four coupled eigenvalue equations. The details are presented in appendix\ref{App-3}. Solving the equations numerically, mass spectra and radial wave functions can be obtained. Accordingly, the normalization condition \eref{E-Norm-D1} can be expressed in terms of radial wave functions as
\begin{equation}\notag
\begin{aligned}
%&\frac{1}{2\times2M}\int \frac{\up{d}^3 q_\perp}{(2\pi)^3} \bar\varphi_\alpha(P,q_\perp,r) \gamma^0 2 \left( \alpha_1    M\hat H( p_{2\perp})+ \w_q \right)d^{\alpha\beta} \varphi_\beta(P,q_\perp,s)\\
%=&\frac{1}{4M}\int \frac{\up{d}^3 q_\perp}{(2\pi)^3}\up{Tr}~ u(r)\bar u(r) \bar A_\alpha  \gamma^0 2 \left( \alpha_1  M\hat H( p_{2\perp})+ \w_q \right) d^{\alpha\beta} A_\beta \\
%\int \frac{\up{d}^3 q_\perp}{(2\pi)^3} &2 \left\{ c_{3}\left[ \w_q(g_1^2+g_2^2)+\alpha_1 M \frac{m_2}{\w_2}(g_1^2-g_2^2)- 2\alpha_1 M\frac{q}{\w_2} g_1g_2\right] \right.\\
%&+c_{1}\left[3\w_q(g_3^2+g_4^2)- 3\alpha_1M \frac{m_2}{\w_2}(g_3^2-g_4^2)+ 2\alpha_1 M \frac{q}{\w_2} g_1g_2 \right. \\
%%&\left.-2\alpha_1M\frac{q}{\w_2}(g_1g_3-g_2g_4) - 2\alpha_1M\frac{m_2}{\w_2}(g_1g_4+g_2g_3)-2\w_q(g_1g_4-g_2g_3)\right]=1,\\
%&\left. \left.-2\alpha_1M\frac{q}{\w_2}(g_1g_3-g_2g_4) -2\left(\alpha_1M\frac{m_2}{\w_2}+\w_q\right)g_1g_3-2\left(\alpha_1M\frac{m_2}{\w_2}-\w_q\right)g_2g_4 \right] \right\}=1,
&\int \frac{\up{d}^3 q_\perp}{(2\pi)^3}  \left[ 2\left(\w_q+ \alpha_1M \frac{m_2}{\w_2} \right)(3c_3g_1^2 - 2c_1g_1g_4+c_1g_4^2) \right. \\
&\left.+2\left( \w_q - \alpha_1M \frac{m_2}{\w_2} \right) (2c_1g_2g_3+c_1g_3^2+3c_3g_2^2)
+4\alpha_1M\frac{q}{\w_2}(c_1g_2g_4+c_1g_3g_4-c_1 g_1g_3-3c_3g_1g_2) \right]=1 \,,
\end{aligned}
\end{equation}
where $c_1 = 1-\frac{\vec q\,^2}{\w_1^2}$ and $c_3 = 1-\frac{\vec q\,^2}{3\w_1^2}$; and the spinor summation $\sum_r u(r)\bar u(r)=\slashchar{P}+M$ is used. 

Similarly, for $J^P=\frac{3}{2}^+$ baryons, the Salpeter wave function can be constructed on the basis of the Rarita-Schwinger spinor $u_\alpha(P,r)$ as
\begin{equation}\label{E-wave-d1-3H}
\begin{aligned}
\varphi_\alpha(P,q_\perp,r)&=\left(t_1+t_2\frac{{\sl q}_\perp}{q}\right)u_\alpha(P,r)+ \left(t_3+t_4\frac{{\sl q}_\perp}{q} \right)\xi_{3\alpha}(P)u_{\hat q_\perp}(P,r)+ \left(t_5+t_6\frac{{\sl q}_\perp}{q}\right)\hat q_{\perp\alpha} u_{\hat q_\perp}(P,r)\\
&=A_{\alpha\beta} u^\beta(P,r)\,,
\end{aligned}
\end{equation}
where we have defined tensor $A_{\alpha\beta}$ for later convenience; $\xi_{3\alpha}(P)=(\gamma_\alpha-\frac{ P_\alpha}{M})$; and $t_i~(i=1,\cdots,6)$ are radial wave functions depending on $q$. According to the Rarita-Schwinger equation, we have $P_\alpha \varphi^\alpha=0$.
%Again, defining a tensor $A_{\alpha\beta}$ as
%\begin{equation}\label{E-tensor-d1-3H}
%A_{\alpha\beta} \equiv \left(t_1+t_2\frac{{\sl q}_\perp}{q} \right)g_{\alpha\beta}+\left(t_3+t_4\frac{{\sl q}_\perp}{q} \right)\xi_{\alpha}\hat{q}_{\perp\beta} + \left(t_5+t_6\frac{{\sl q}_\perp}{q} \right)\hat{q}_{\perp\alpha} \hat{q}_{\perp\beta},
%\end{equation}
%we can have
%\begin{equation}\label{E-wave2-d1-3H}
%\varphi_\alpha(P,q_\perp,r)=A_{\alpha\beta} u^\beta(P,r).
%\end{equation}
Note that the full Salpeter wave function has six independent terms, compared with Ref.~\cite{WengMH2011}, where only the first two of them, i.e., $t_1$ and $t_2$, are considered. The decomposition in terms of spherical harmonics is similar with the case in $J^P=\frac{1}{2}^+$ and we would not repeat the details. It is evident that $t_{3}$, $t_4$, $t_5$, and $t_6$ correspond to the $P$, $^2D$, $^4D$, and $F$ partial waves respectively. Lacking of them means that only the largest $S$-wave and part of the $P$-wave components are considered. Notice that the $D(F)$-wave would always be mixed with the $S(P)$-wave components, namely, the $t_{4(5)}$ would always contribute to the $S$-wave besides the $D$-wave components, and $t_6$ would also contribute to the $P$-wave besides the $F$-wave components. Further analysis of their importance will be presented later.

Now we obtain the following mass eigenvalue equation
\begin{equation} \label{E-eigen-D1-3H}
MA_{\alpha\beta}u^{\beta}(P,r)=(\w_1+\w_2)\hat H( p_{2\perp})A_{\alpha\beta} u^\beta(P,r) + \vartheta_{\alpha\beta}\hat{H}(p_{2\perp}) \gamma^0 \int \frac{\up d^3 k_\perp}{(2\pi)^3} K(k_\perp-q_\perp) A^{\beta\gamma}u_\gamma(P,r) \,,
\end{equation}
which is equivalent to six coupled eigenvalue equations similar with that for the $\frac{1}{2}^+$ case, and we will not copy details again. The normalization condition can be written as
\begin{equation}\notag
\begin{aligned}
%&\frac{1}{(2\dx \frac{3}{2}+1)\times2M}\int \frac{\up{d}^3 q_\perp}{(2\pi)^3} \bar\varphi_\alpha(P,q_\perp,r) 2 \left(\alpha_1    M\hat H( p_{2\perp})+\w_q \right) d^{\alpha\beta} \varphi_\beta(P,q_\perp,r)\\
%=&\frac{1}{8M}\int \frac{\up{d}^3 q_\perp}{(2\pi)^3}\up{Tr}~ u^\mu(P,r)\bar u^\nu(P,r)\bar A_{\alpha\nu} 2\left(\alpha_1    M\hat H( p_{2\perp})+\w_q \right) A_{\beta\mu} d^{\alpha\beta}\\
&\int \frac{\up{d}^3 q_\perp}{(2\pi)^3} \frac{2}{3} \left \{
\left[3 c_3( t_1^2+t_4^2)-4 c_2t_1 t_4 -2 c_1 ( t_1 t_5- t_4 t_5)+c_1 t_5^2\right]\left(\omega _q+\alpha _1 M \frac{m_2}{\w_2}\right) \right.\\
&+ \left[3c_3( t_2^2+t_3^2)+4c_2 t_2t_3-2c_1 (t_2t_6+t_3t_6)+c_1 t_6^2   \right]\left(\omega _q-\alpha _1 M \frac{m_2}{\w_2}\right) \\
&\left. -2 \alpha _1 M \frac{q }{\omega _2}\left[3 c_3 (t_1 t_2-t_3t_4)+2 c_2 (t_1t_3- t_2 t_4) -c_1(t_1t_6+t_2t_5+t_3t_5 -t_4t_6-t_5t_6)\right]
\right\}=1,
\end{aligned}
\end{equation}
where $c_2=1-\frac{\vec q\,^2}{2\w_1^2}$, and the following completeness relation of the Rarita-Schwinger spinor \cite{Behrends1957} has been used:
\begin{equation}\notag
u^\alpha(P,r) \bar u^\beta(P,r)=(\sl P+M)\left[ -g^{\alpha\beta}+\frac{1}{3}\gamma^\alpha\gamma^\beta - \frac{P^\alpha\gamma^\beta-P^\beta\gamma^\alpha}{3M}+\frac{2P^\alpha P^\beta}{3M^2} \right].
\end{equation}

\section{Numerical results and discussions}\label{Sec-4}
For solving the Salpeter equation of baryons, we need to specify the parameter $V_0$ in the potential for the diquark core and the light quark. Unlike doubly heavy mesons, experimental data of doubly heavy baryons is very limited. So one cannot fix $V_0$ by fitting experimental data. Alternatively, here we compute $V_0$ by taking the spin-weighted average of the corresponding mesons' $V_0$ (see details in appendix\ref{App-5}).  The fixed values of $V_0$ are listed in \autoref{Tab-V0}. Note that all the parameters in the model are now fixed by the meson sector.

\begin{table}[h!]
\caption{The relevant parameter $(-V_0)$ (in GeV) determined by spin-weighted average method.}
\label{Tab-V0}
\vspace{0.2em}\centering
\begin{tabular}{ c|ccccccccccccc }
\toprule[2pt]
%\hline
%\hline
%\multirow{4}{*}{I} & & $\Xi_{cc}$		& $\Omega^+_{cc}$	&$\Xi_{cb}$ & $\Omega_{cb}$ & $\Xi_{bb}$		& $\Omega_{bb}$	&  \\
%& & $0.460$	  &$0.425$	  &$0.375$	  &$0.385$	  &$0.329$	  &$0.270$	  &	\\
%%\midrule[1.5pt]
%& & $\Xi_{cc}^{*}$		& $\Omega_{cc}^{*}$		& $\Xi^*_{cb}$	&$\Omega_{cb}^{*}$	& $\Xi^{*}_{bb}$		& $\Omega^*_{bb}$	& \\
%& & $0.381$	  &$0.346$	  &$0.350$	  &$0.355$	  &$0.330$	  &$0.274$	  &	\\
%\midrule[1pt]
$\frac{1}{2}^+$ baryons & $\Xi_{cc}^{++}$		& $\Xi_{cc}^+$		& $\Omega^+_{cc}$	&$\Xi_{cb}^{+}$	& $\Xi_{cb}$		& $\Omega_{cb}$	&$\Xi_{bb}$		& $\Omega_{bb}^-$\\
$-V_0$  & $0.478$	  &$0.476$	  &$0.454$	  &$0.404$	  &$0.403$	  &$0.382$	  &$0.330$	  &$0.310$	\\
\midrule[1.5pt]
%\hline
$\frac{3}{2}^+$ baryons & $\Xi_{cc}^{*++}$		& $\Xi_{cc}^{*+}$		& $\Omega^*_{cc}$	&$\Xi_{cb}^{*+}$	& $\Xi^{*0}_{cb}$		& $\Omega^*_{cb}$	&$\Xi^*_{bb}$		& $\Omega_{bb}^{*-}$\\
$-V_0$  & $0.378$	  &$0.376$	  &$0.352$	  &$0.337$	  &$0.336$	  &$0.313$	  &$0.296$	  &$0.275$	\\
\bottomrule[2pt]
%\hline
%\hline
\end{tabular}
\end{table}

In this work, specifically, we calculate $\Xi^{(*)}_{cc}$, $\Omega^{(*)}_{cc}$, $\Xi^{(*)}_{cb}$, $\Omega^{(*)}_{cb}$, $\Xi^{(*)}_{bb}$, and $\Omega^{(*)}_{bb}$ for $J^P=\frac{1}{2}^+(\frac{3}{2}^+)$ doubly heavy baryons.
In the approach of relativistic Bethe-Salpeter equation, the total angular momentum $J$ and the space parity $P$ of baryons are good quantum numbers. Here, to reflect the dominant characteristic, we still label baryon states by the nonrelativistic notations as $n_\up{B}{^{2s_\up{B}+1}L_{J}}(n_\up{D}L_\up{D})$: $n_\up{D}$ denotes the radial quantum number of the heavy diquark core; $L_\up{D}$, the orbital angular momentum of the diquark; $n_\up{B}$, the radial quantum number of the baryon; $L$, the quantum number of the diquark-quark orbital angular momentum; $(2s_\up{B}+1)$, the multiplicity of the baryon spin $s_\up{B}$; and $J$, the total angular momentum of the baryon. Consequently, $\frac{1}{2}^+$ baryons can have states: $ {^2S_{1/2}}$ or $ {^4D_{1/2}}$, or their mixing; and $\frac{3}{2}^+$ baryons: $ {^4S_{3/2}}$, $ {^4D_{3/2}}$, $ {^2D_{3/2}}$, or their mixing.

\begin{figure}[h!]
\centering
\subfigure[$n=1$: mainly the $1S$ components]   {\includegraphics[width=0.42\textwidth]{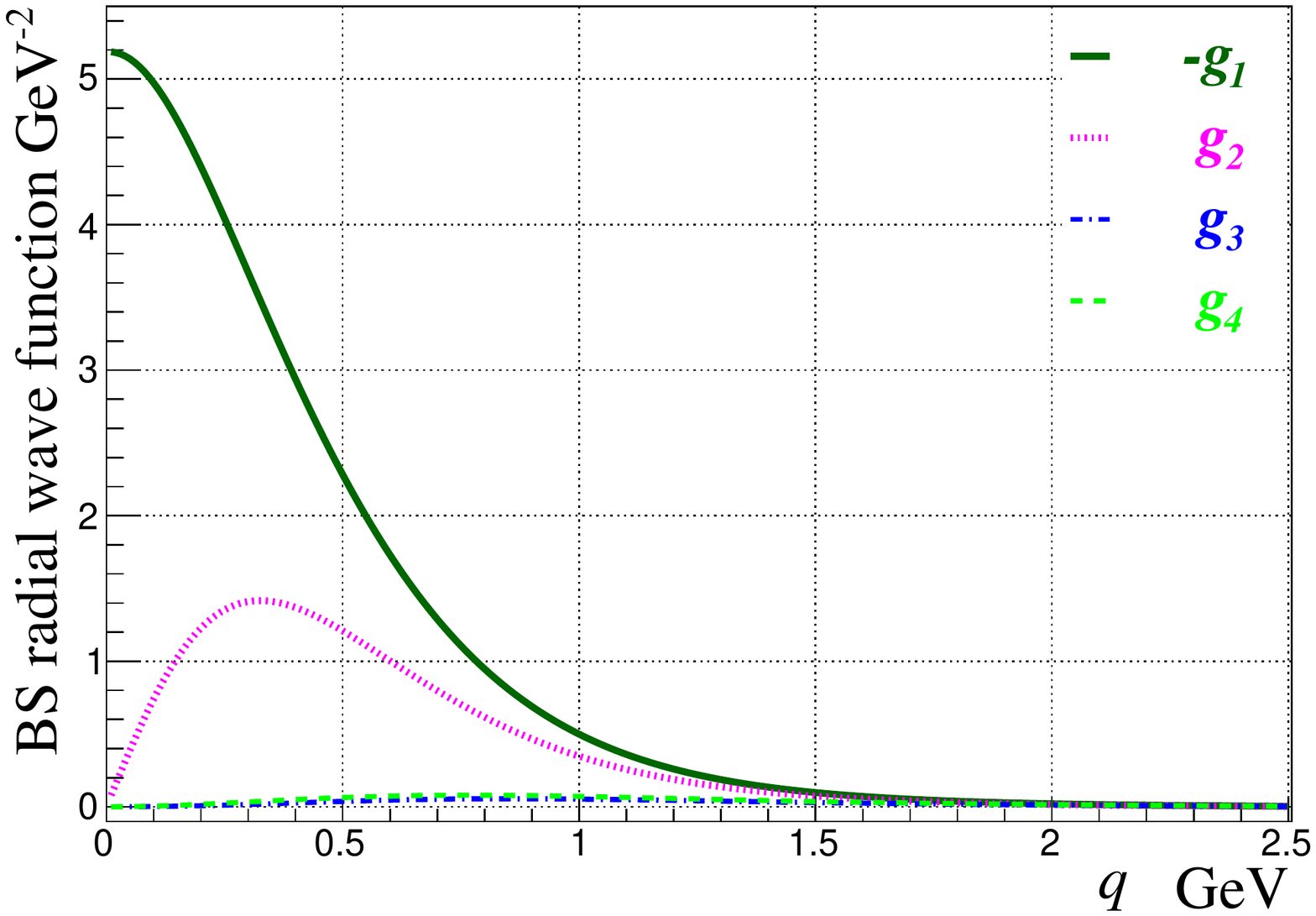} \label{Fig-Xicc-n1}}
\subfigure[$n=2$: $2S$ mixed with $1D$]   {\includegraphics[width=0.42\textwidth]{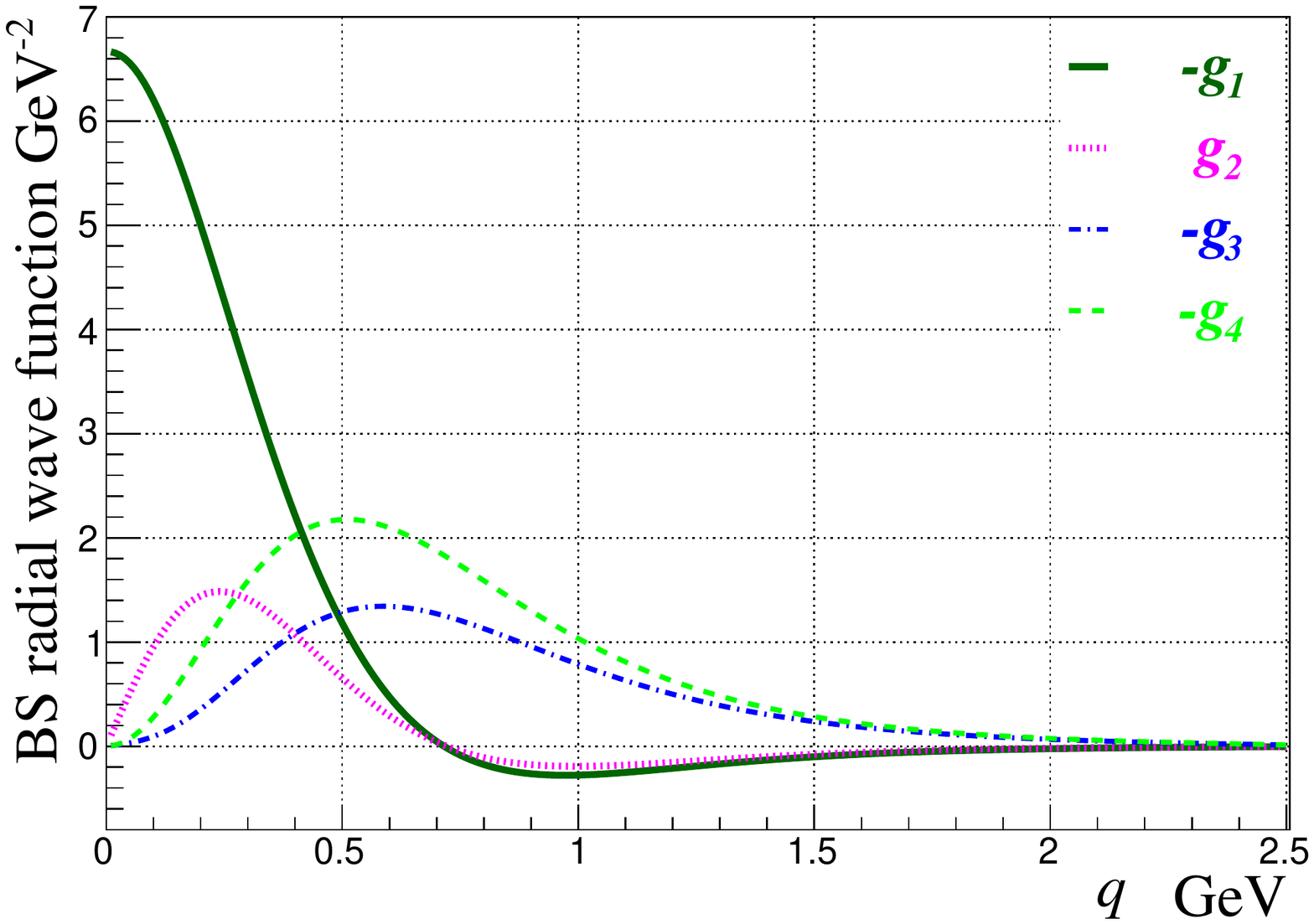} \label{Fig-Xicc-n2}}
\subfigure[$n=3$: $1D$ mixed with $2S$]   {\includegraphics[width=0.42\textwidth]{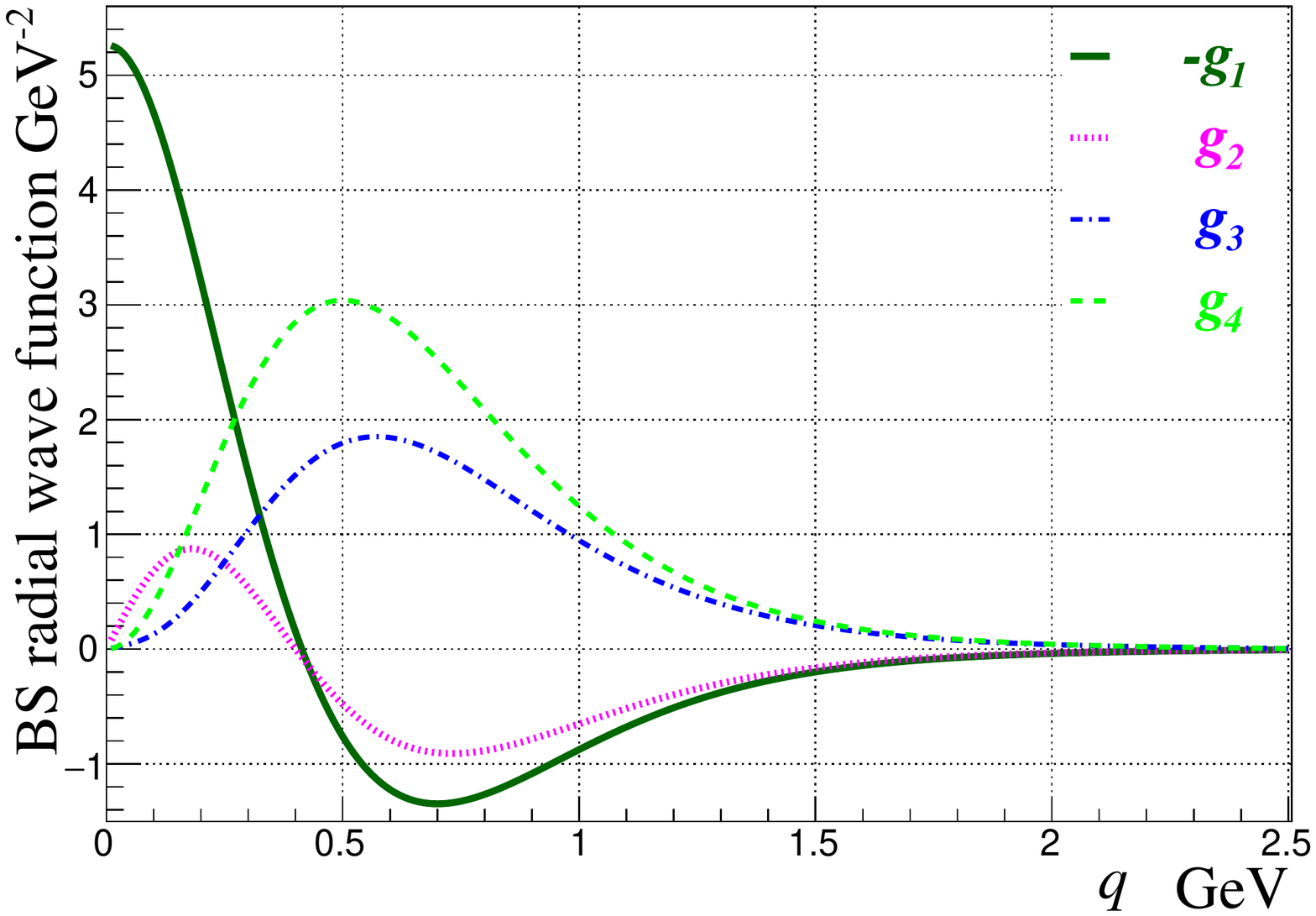} \label{Fig-Xicc-n3}}
\subfigure[$n=4$: $3S$ mixed with $2D$]   {\includegraphics[width=0.42\textwidth]{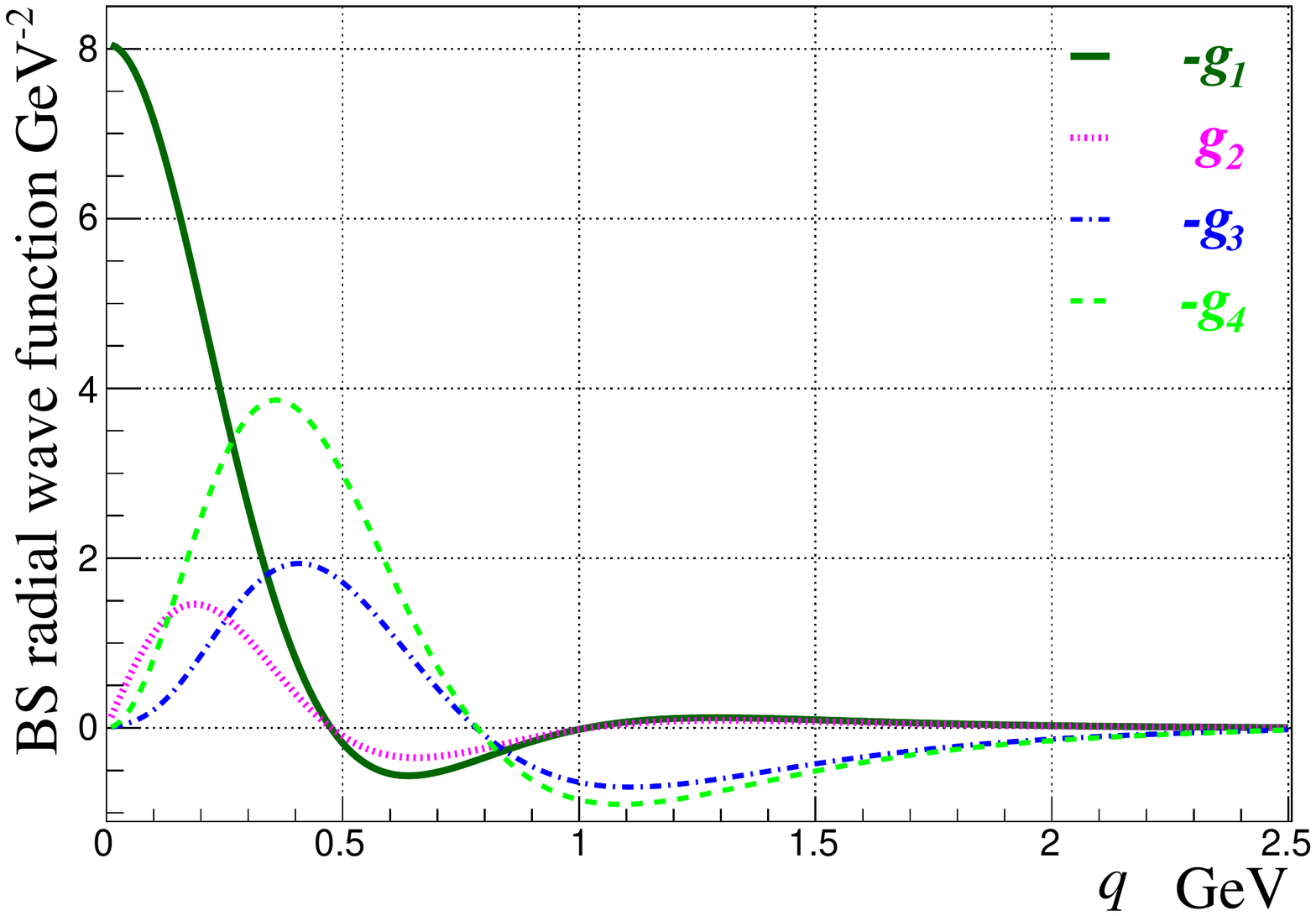} \label{Fig-Xicc-n4}}
\caption{BS radial wave functions of the $\Xi_{cc}^{++}$ with the energy level $n=1,\cdots,4$; $g_{1}$ and $g_{4}$ correspond to the $^2\hm S$ and $^4\hm D$ components respectively, and $g_{2(3)}$ corresponds to the $P$-wave;  $n_\up{B}$ is one more than the number of the node; and almost every state contains the $S$-, $P$-, and $D$-wave components.} \label{Fig-Xicc}
\end{figure}

\begin{table}[h!]
\caption{Mass spectra for $J^P=\frac{1}{2}^+$ doubly heavy baryons in units of MeV.  Symbols used to label the baryon states are: $n_\up{D}$ denotes the radial quantum number of the doubly heavy diquark core inside the baryon; $L_\up{D}$, the orbital angular momentum of the diquark; $n_\up{B}$, the radial number of the baryon; $(2s_\up{B}+1)$, the baryon spin multiplicity; $L$, the orbital angular momentum quantum number between the diquark core and the light quark; and finally $J$, the total baryon angular momentum.}
\label{Tab-Mass-1-2}
\vspace{0.2em}\centering\setlength{\tabcolsep}{3pt}
\begin{tabular}{ lcccccccccccc }
\toprule[2pt]
$n$		&$n_\up{B}{^{2s_\up{B}+1}L_J}(n_\up{D}L_\up{D})$		& $\Xi_{cc}^{++}$		& $\Xi_{cc}^+$		& $\Omega^+_{cc}$	&$\Xi_{cb}^{+}$	& $\Xi_{cb}$		& $\Omega_{cb}$	&$\Xi_{bb}$	& $\Xi_{bb}^-$		& $\Omega_{bb}^-$\\
\midrule[1.5pt]
1  &$1^2S_{1/2}(1S)$  &$ 3601^{-28}_{+28}$  &$ 3606^{-28}_{+28}$  &$ 3710^{-28}_{+27}$  &$ 6931^{-28}_{+27}$  &$ 6934^{-28}_{+27}$  &$ 7033^{-26}_{+24}$  &$10182^{-25}_{+25}$  &$10184^{-25}_{+25}$  &$10276^{-23}_{+22}$ \\ 
2  &$2^2S_{1/2}(1S)$  &$ 4122^{-38}_{+38}$  &$ 4128^{-40}_{+38}$  &$ 4247^{-38}_{+38}$  &$ 7446^{-37}_{+38}$  &$ 7450^{-39}_{+38}$  &$ 7560^{-36}_{+35}$  &$10708^{-35}_{+35}$  &$10710^{-36}_{+34}$  &$10816^{-34}_{+32}$ \\ 
3  &$1^4D_{1/2}(1S)$  &$ 4151^{-38}_{+39}$  &$ 4157^{-38}_{+38}$  &$ 4289^{-37}_{+36}$  &$ 7463^{-35}_{+35}$  &$ 7467^{-34}_{+35}$  &$ 7598^{-33}_{+33}$  &$10732^{-32}_{+33}$  &$10735^{-32}_{+33}$  &$10863^{-32}_{+32}$ \\ 
4  &$3^2S_{1/2}(1S)$  &$ 4504^{-54}_{+54}$  &$ 4510^{-54}_{+54}$  &$ 4632^{-53}_{+52}$  &$ 7818^{-51}_{+52}$  &$ 7822^{-52}_{+51}$  &$ 7935^{-49}_{+50}$  &$11084^{-49}_{+48}$  &$11086^{-50}_{+48}$  &$11196^{-48}_{+47}$ \\
1  &$1^2S_{1/2}(2S)$  &$ 4136^{-36}_{+35}$  &$ 4141^{-35}_{+35}$  &$ 4261^{-33}_{+33}$  &$ 7417^{-32}_{+32}$  &$ 7420^{-33}_{+33}$  &$ 7531^{-30}_{+32}$  &$10618^{-30}_{+28}$  &$10620^{-29}_{+28}$  &$10724^{-28}_{+26}$ \\ 
1  &$1^2S_{1/2}(1D)$  &$ 4140^{-33}_{+33}$  &$ 4145^{-33}_{+33}$  &$ 4262^{-32}_{+32}$  &$ 7438^{-32}_{+31}$  &$ 7441^{-32}_{+31}$  &$ 7550^{-30}_{+30}$  &$10660^{-28}_{+28}$  &$10661^{-28}_{+28}$  &$10763^{-26}_{+26}$ \\
\bottomrule[2pt]
\end{tabular}
\end{table}
The mass spectra for $J^P=\frac{1}{2}^+$ doubly heavy baryons with flavors $(ccq)$, $(bcq)$ and $(bbq)$ are presented in \autoref{Tab-Mass-1-2}, including the diquark cores at the excited $2S$ and $1D$ states. Notice the baryon masses of $1^2S_{1/2}(1D)$ are comparable with that of the $1^4D_{1/2}(1S)$ state, namely, the different $D$-wave orbital angular momentum contribute similarly to the total baryon mass.  The wave functions of $\Xi_{cc}^{++}$, as an example, are presented in \autoref{Fig-Xicc}. From the curves in the figures one may see the correspondence of the radial quantum number $n_\up{B}$ and the node number of the wave functions clearly; $g_{1}$ and $g_{4}$ correspond to the $^2\hm S$ and $^4\hm D$ partial wave components respectively; $g_2$ and $g_3$ represent the slight $P$-wave components with different radial number. Comparing the weights of the partial wave components, one can realize that the ground state $n=1$ is dominated by the $S$-wave and has a negligible $D$-wave, whereas, the excited states $n=2,3,4\dots$ have sizable $D$-wave components. It means that for excited states, to ignore the components $g_{3,4}$ would destroy the results. Therefore, as mentioned in the previous section, the full structures of the wave functions should be well considered.

\begin{table}[h!]
\caption{Mass spectra (in MeV) for the $\frac{3}{2}^+$ doubly heavy baryons.}
%The symbols to denote the baryon states are the same with that in \autoref{Tab-Mass-1-2}.}
\label{Tab-Mass-3-2}
\vspace{0.2em}\centering \setlength{\tabcolsep}{3pt}
\begin{tabular}{ lcccccccccccc }
\toprule[2pt]
$n$		&$n_\up{B}{^{2s_\up{B}+1}\! L_J}(n_\up{D}L_\up{D})$	& $\Xi_{cc}^{*++}$		& $\Xi_{cc}^{*+}$		& $\Omega^*_{cc}$		&$\Xi_{cb}^{*+}$	& $\Xi^*_{cb}$		& $\Omega^*_{cb}$		&$\Xi_{bb}^{*}$	& $\Xi^{*-}_{bb}$		& $\Omega^{*-}_{bb}$\\
\midrule[1.5pt]
1  &$1^4S_{3/2}(1S)$  &$ 3703^{-28}_{+28}$  &$ 3706^{-28}_{+28}$  &$ 3814^{-27}_{+27}$  &$ 6997^{-27}_{+27}$  &$ 7000^{-27}_{+25}$  &$ 7101^{-25}_{+25}$  &$10214^{-25}_{+25}$  &$10216^{-24}_{+25}$  &$10309^{-23}_{+24}$ \\ 
2  &$2^4S_{3/2}(1S)$  &$ 4232^{-40}_{+40}$  &$ 4237^{-41}_{+38}$  &$ 4356^{-38}_{+38}$  &$ 7512^{-39}_{+36}$  &$ 7515^{-38}_{+36}$  &$ 7627^{-36}_{+33}$  &$10738^{-35}_{+35}$  &$10740^{-35}_{+35}$  &$10847^{-32}_{+33}$ \\ 
3  &$1^2D_{3/2}(1S)$  &$ 4252^{-37}_{+36}$  &$ 4257^{-37}_{+36}$  &$ 4395^{-35}_{+36}$  &$ 7531^{-35}_{+33}$  &$ 7535^{-36}_{+32}$  &$ 7668^{-35}_{+32}$  &$10766^{-32}_{+32}$  &$10769^{-32}_{+32}$  &$10898^{-32}_{+32}$ \\ 
4  &$1^4D_{3/2}(1S)$  &$ 4421^{-46}_{+45}$  &$ 4424^{-46}_{+45}$  &$ 4520^{-43}_{+43}$  &$ 7699^{-43}_{+40}$  &$ 7701^{-43}_{+40}$  &$ 7791^{-41}_{+37}$  &$10941^{-40}_{+38}$  &$10943^{-39}_{+38}$  &$11028^{-37}_{+36}$ \\ 
1  &$1^4S_{3/2}(2S)$  &$ 4237^{-36}_{+35}$  &$ 4241^{-35}_{+35}$  &$ 4365^{-33}_{+33}$  &$ 7484^{-33}_{+32}$  &$ 7487^{-33}_{+32}$  &$ 7601^{-32}_{+30}$  &$10651^{-28}_{+28}$  &$10653^{-30}_{+28}$  &$10759^{-28}_{+28}$ \\
1  &$1^4S_{3/2}(1D)$  &$ 4241^{-33}_{+33}$  &$ 4245^{-33}_{+32}$  &$ 4366^{-32}_{+32}$  &$ 7504^{-30}_{+31}$  &$ 7507^{-30}_{+31}$  &$ 7619^{-29}_{+30}$  &$10693^{-28}_{+28}$  &$10695^{-28}_{+28}$  &$10798^{-26}_{+26}$ \\
\bottomrule[2pt]
\end{tabular}
\end{table}

\begin{figure}[h!]
\centering
\subfigure[$n=1$: mainly the $1S$]   {\includegraphics[width=0.42\textwidth]{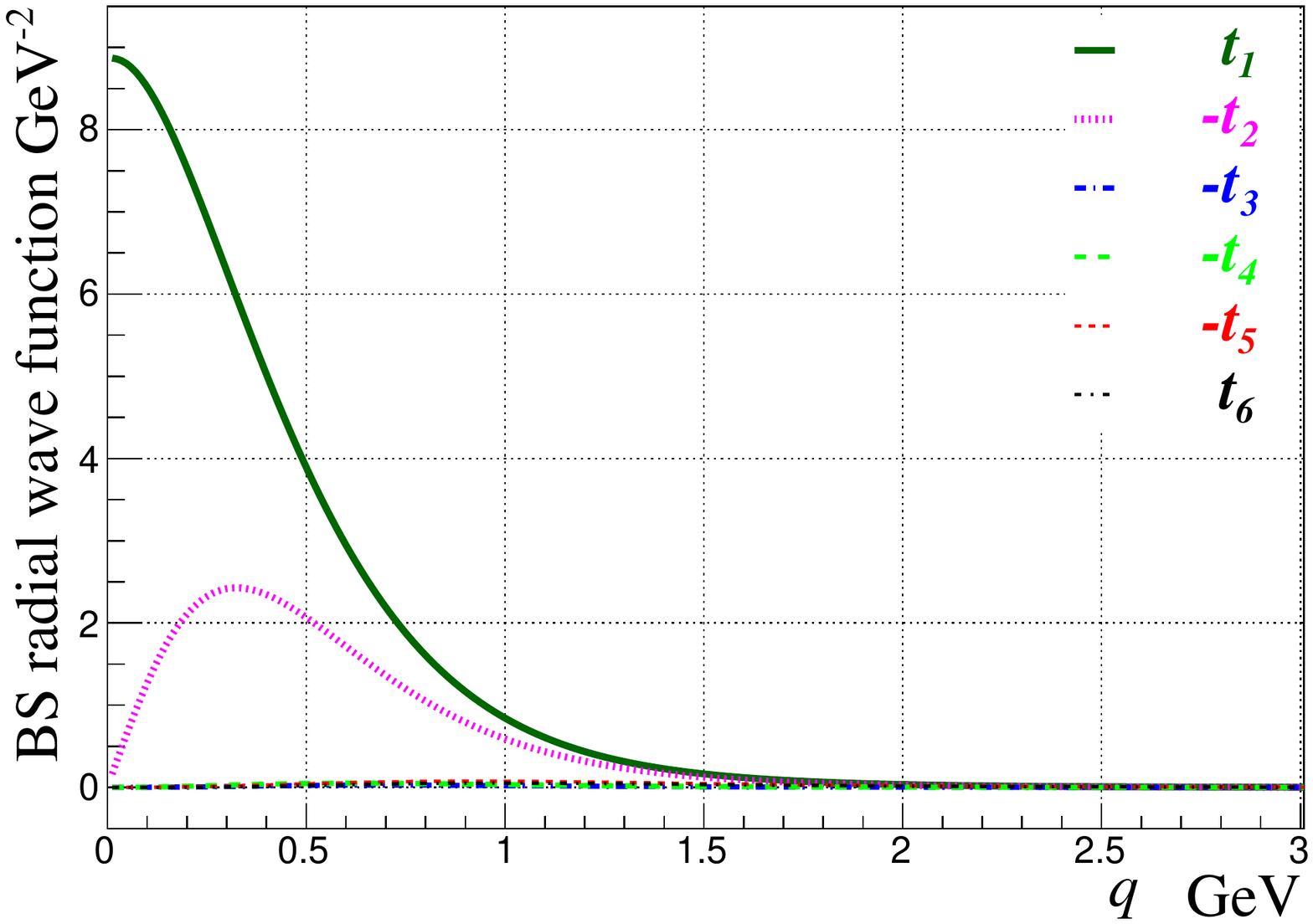} \label{Fig-Xiccs-n1}}
\subfigure[$n=2$: the $2S$ with slight $1^2D$]   {\includegraphics[width=0.42\textwidth]{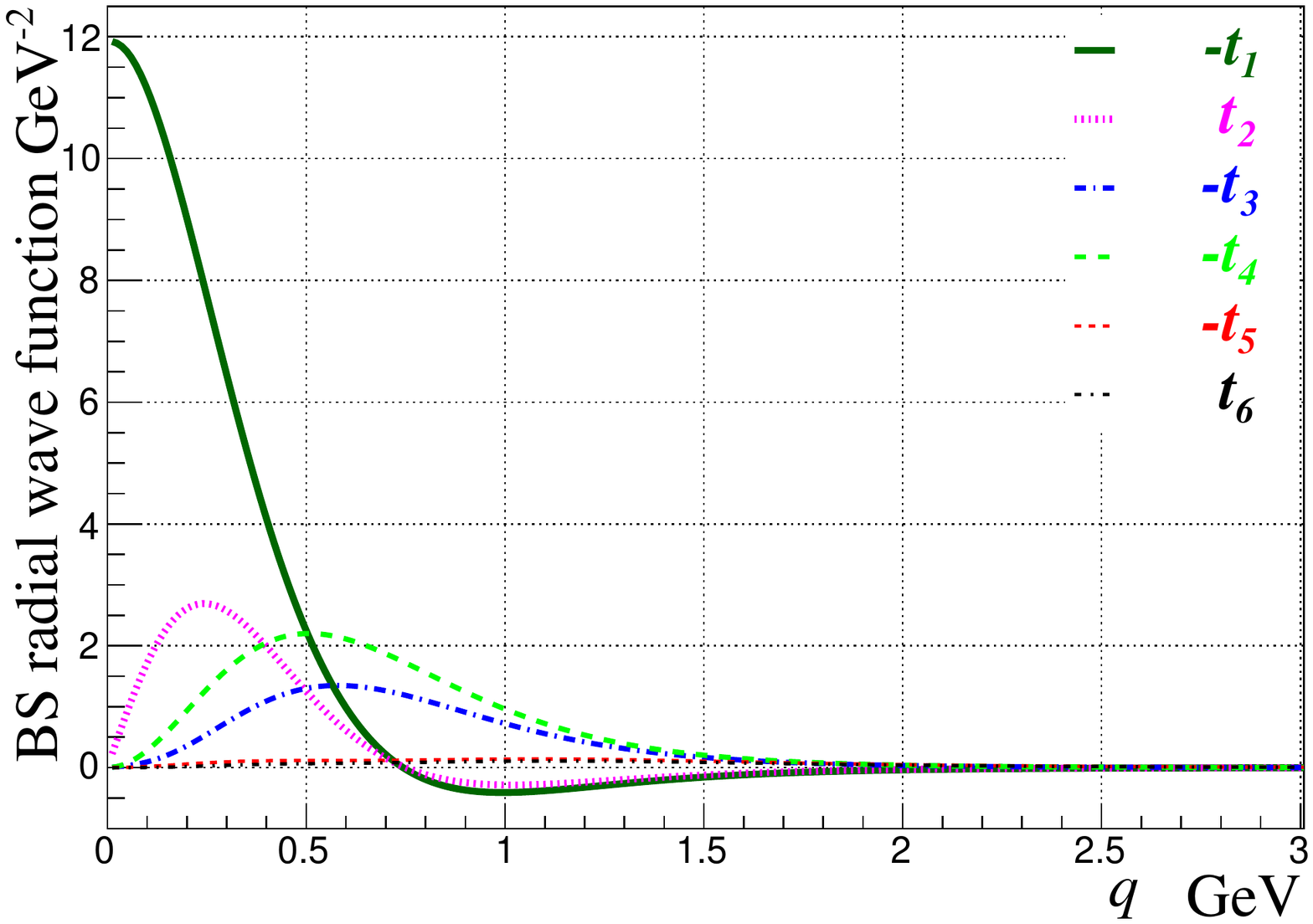} \label{Fig-Xiccs-n2}}
\subfigure[$n=3$: the $1^2D$ mixed with $2S$]   {\includegraphics[width=0.42\textwidth]{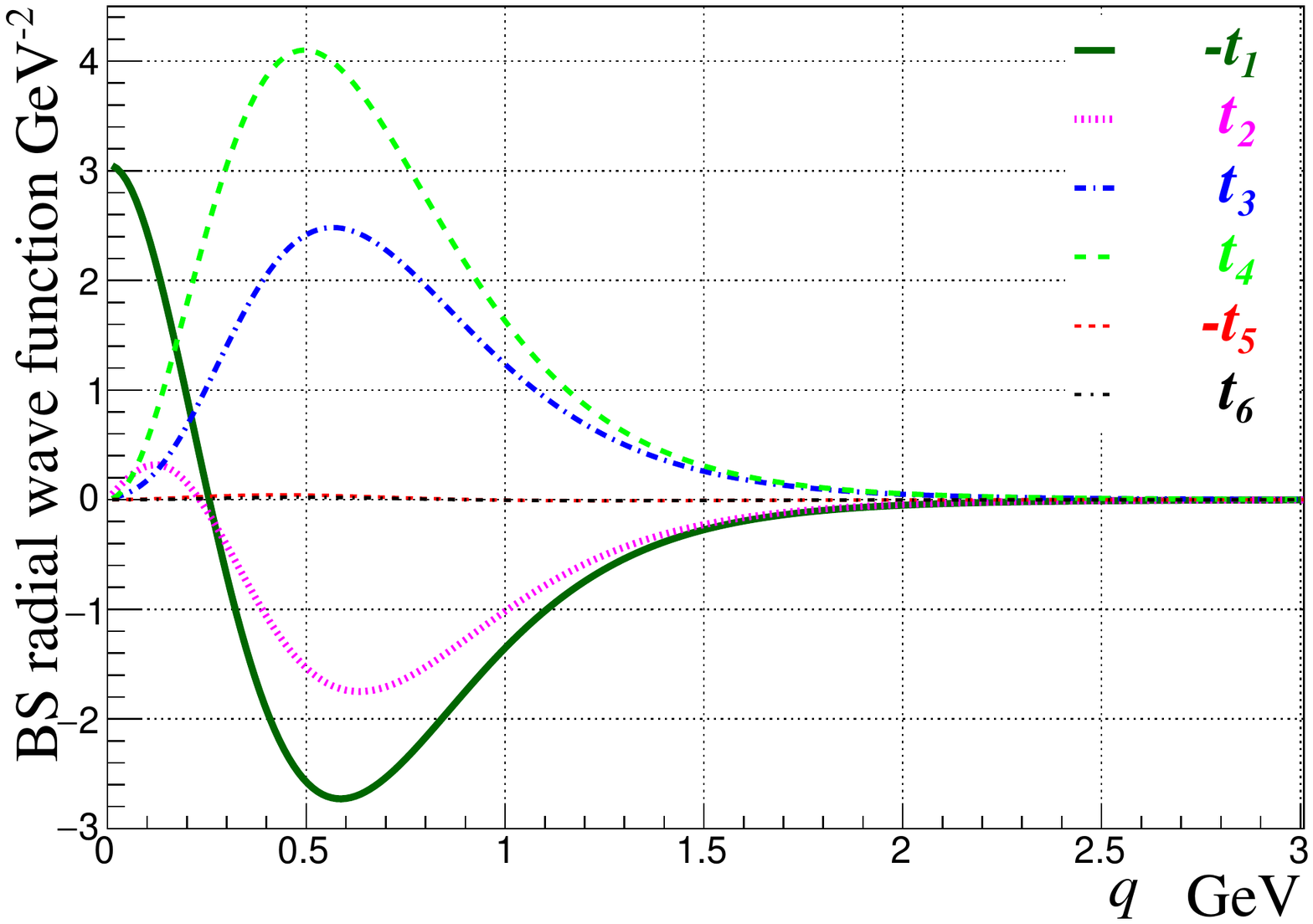} \label{Fig-Xiccs-n3}}
\subfigure[$n=4$: $1^4D$ with slight $2S$ and $1^2D$ mixed in]   {\includegraphics[width=0.42\textwidth]{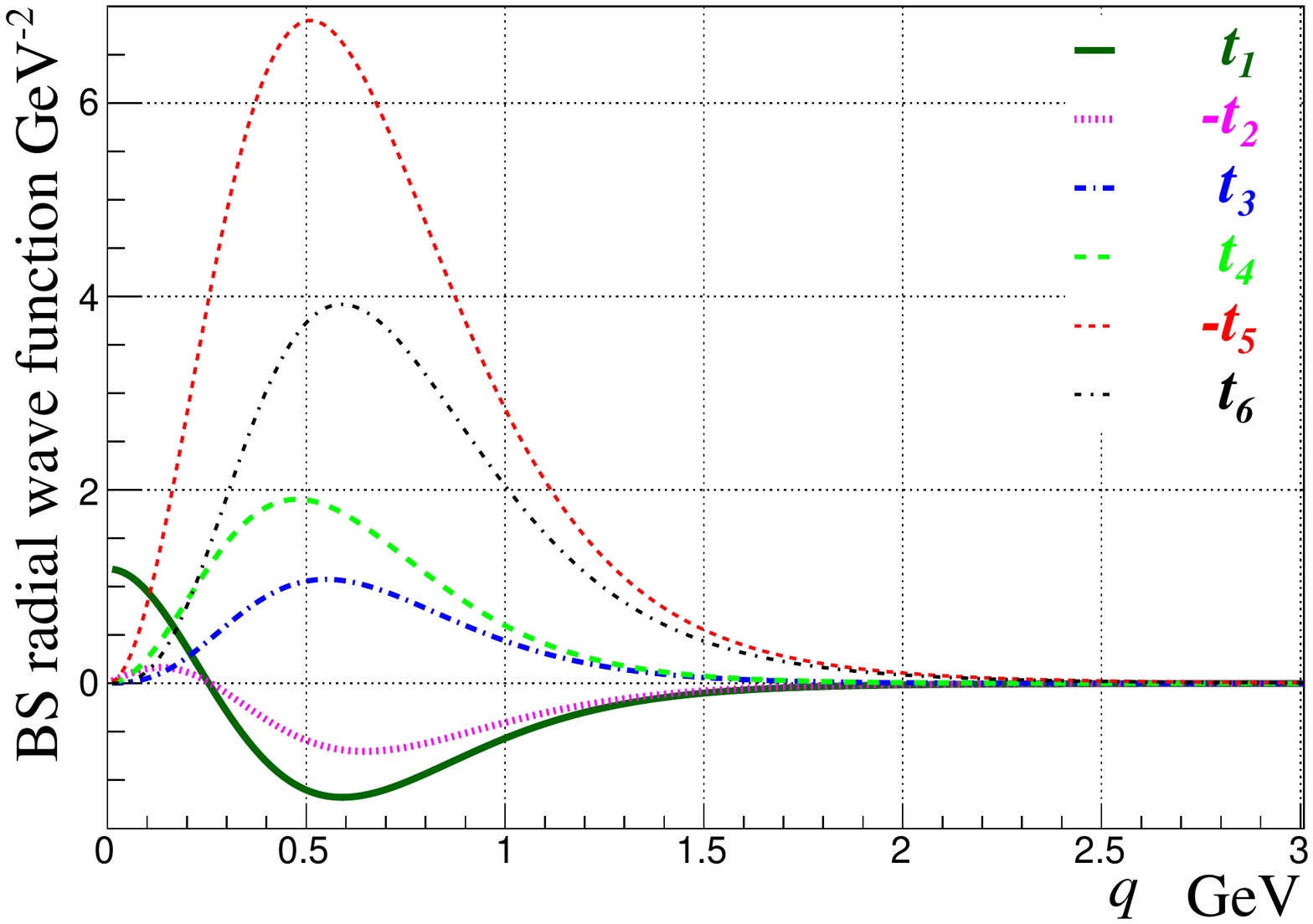} \label{Fig-Xiccs-n4}}
\caption{BS radial wave functions of the $\Xi_{cc}^{*}$ with the energy level $n=1,\cdots,4$; $t_{1}$, $t_{4}$ and $t_{5}$ correspond to the $^4\hm S$, $^2\hm D$ and $^4\hspace{-0.1em}D$ components respectively, $t_{2(3)}$ and $t_6$ correspond to the $P$ and $F$-wave respectively;  and also almost every baryon state contains all the $^4\hm S$, $^2\hm D$ and $^4\hspace{-0.1em}D$ components.} \label{Fig-Xicc-star}
\end{figure}

The mass spectra for $J^P=\frac{3}{2}^+$ doubly heavy baryons are presented in \autoref{Tab-Mass-3-2}, and the radial wave functions of $\Xi_{cc}^*$ are shown in \autoref{Fig-Xicc-star} as an example.
For $J^P=\frac{3}{2}^+$ baryons, besides the ${^{4}S_{3/2}}$ partial waves, they also contain two different $D$-waves, i.e., ${^4 D_{3/2}}$ and  ${^2D_{3/2}}$, which belongs to the quartet $(\frac{7}{2},\frac{5}{2},\frac{3}{2},\frac{1}{2})^+$ and the doublet $(\frac{5}{2},\frac{3}{2})^+$, respectively. The baryon masses with the excited $2S$ and $1D$ diquarks are listed in the last two rows to make an comparison with the mass spectra with ground diquarks. Note that the masses of $1^4S_{3/2}(1D)$ states are close to those of $1^2D_{3/2}(1S)$ states, but much lower than those of the $1^4D_{3/2}(1S)$ states.
The similar analysis on the wave functions shows that almost every state contains all the possible $S$-, $P$-, $D$-, and $F$-wave components. And the analysis on the weight of the partial waves suggests that the $n=1,2$ states are mainly characterized by $1\hp{^{4}S}$ and $2\hp{^{4}S}$ components, respectively, with $D$-wave mixed slightly. However, for the $n=3,4$ states, the $1^2D$ and $1^4D$ components become dominant, respectively. Again, these features emphasize the importance to involve the full structures of the wave functions. 

To see the sensitivity of the mass spectra on the model parameters, we calculate the theoretical uncertainties by varying the parameters $\lambda$, and $a_{1(2)}$ in the quark-diquark bound potential by $\pm5\%$ simultaneously, and then searching the parameter space to find the maximum deviation. The obtained theoretical errors are also listed in above \autoref{Tab-Mass-1-2} and \autoref{Tab-Mass-3-2}, where the relative errors induced from $\lambda$ and $a_{1(2)}$ are about $1\%$ roughly. The dependence on the constituent quark mass is almost linear and not included in the error analysis.

Finally, we give a comparison between the predicted spectra of ground states and those in literatures, which is collected in \autoref{Tab-Mass-C}. We notice that the mass spectra agree well with the lattice QCD calculations in Refs.\,\cite{Brown2014,Mathur2018,Mathur2019}, especially for $\Xi^{(*)}_{cc}$, $\Omega^{*}_{cc}$, $\Omega_{cb}$, and $\Omega^{(*)}_{bb}$.  The average magnitude of errors in predictions of \autoref{Tab-Mass-C} with that in Ref.\,\cite{Brown2014} is about 18 MeV.

%---------------------------------------------------------------------------
\begin{table}[h!]
\caption{Comparisons of the predictions for the ground state masses (in GeV) of the doubly heavy baryons, where the results of the third\,\cite{Brown2014} and forth\,\cite{Mathur2018,Mathur2019} columns are from the lattice QCD calculations. Note that here $\Xi_{bc}$, $\Omega_{bc}$ denote the $\frac{1}{2}^+$ $(bcq)$-baryons with $1^+$ $(bc)$-diquark core only.}
\label{Tab-Mass-C}
\vspace{0.2em}\centering
\begin{tabular}{ cccccccccccc }
\toprule[2pt]
Baryon 			& This		&\cite{Brown2014}	&\cite{Mathur2018,Mathur2019}	&\cite{Karliner2014,Karliner2018}	 &\cite{Ebert2002}	&\cite{Albertus2007}	 &\cite{Giannuzzi2009} &\cite{WengXZ2018}	&\cite{LvQF2017}\\
\midrule[1.5pt]
$\Xi_{cc}$			&3.601	&3.610	&-			&3.627				& 3.620				&3.612	&3.547	&3.633	&3.606\\
$\Xi_{cc}^*$		&3.703	&3.692	&-			&3.690 				&3.727				&3.706	&3.719	&3.696	&3.675\\
$\Omega_{cc}$	&3.710	&3.738	&3.712	&3.692         			&3.778				&3.702	&3.648	&3.732	&3.715\\
$\Omega^*_{cc}$	&3.814	&3.822	&3.788	&3.756 				&3.872				&3.783	&3.770	&3.802	&3.772\\
$\Xi_{cb}$			&6.931	&6.959	&6.945	&6.933 				&6.933				&6.919	&6.904	&6.948	&-\\
$\Xi^*_{cb}$		&6.997	&6.985	&6.989	&6.969 				&6.980				&6.986	&6.936	&6.973	&-\\
$\Omega_{cb}$	&7.033	&7.032	&6.994	&6.984 				&7.088				&6.986	&6.994	&7.047	&-\\
$\Omega^*_{cb}$&7.101	&7.059	&7.056	&- 						&7.130				&7.046	&7.017	&7.066	&-\\
$\Xi_{bb}$		&10.182	&10.143	&-			&10.162 				&10.202				&10.197	&10.185	&10.169	&10.138\\
$\Xi^*_{bb}$		&10.214	&10.178	&-			&10.184 				&10.237				&10.236	&10.216	&10.189	&10.169\\
$\Omega_{bb}$	&10.276	&10.273	&-			&10.208 				&10.359				&10.260	&10.271	&10.259	&10.230\\
$\Omega_{bb}^*$&10.309	&10.308	&-			&- 						&10.389				&10.297	&10.289	&10.268	&10.258\\
\bottomrule[2pt]
\end{tabular}
\end{table}

In summary, within the diquark core picture, we have built a theoretical framework to deal with doubly heavy baryons. 
	Based on the diquark picture, we reduce the three-body problem of the doubly heavy baryons into two two-body bound problems and then resolved by the corresponding BSE in the instantaneous approximation. We obtain the mass spectra and wave functions of $J^P=\frac{1}{2}^+$ and $\frac{3}{2}^+$ doubly heavy baryons with $1^+$ diquark cores. Roughly speaking, the predicted mass spectra for the doubly heavy baryons agree with those in the literature, especially those by lattice QCD computation. Analyzing the wave functions, we found that $\frac{1}{2}^+$ and $\frac{3}{2}^+$ baryons, especially their excited states, involve unneglectable $P$- and $D$-wave components, and even $F$-wave for the latter, which means that the non-relativistic wave functions only consisting of $S$ or $D$ partial waves cannot work well to describe these baryons. In the near future, the obtained wave functions can be used to calculate lifetimes, productions, and decays of doubly heavy baryons, which can also be tested by the coming experiments.

\section*{Acknowledgments}
We thank Hai-Yang Cheng, Xing-Gang Wu, Xu-Chang Zheng, Tian-Hong Wang and Hui-Feng Fu for the helpful discusses and suggestions. This work was supported in part by the National Natural Science Foundation of China (NSFC) under Grant Nos.\,11447601, 11535002, 11575048, 11675239, 11805024,  and 11821505. It's also supported by the Key Research Program of Frontier Sciences, CAS, Grant No.\,QYZDY-SSW-SYS006, the China Postdoctoral Science Foundation under Grant No.\,2018M641487, and the Fundamental Research Funds for the Central Universities under Grant No.\,G2019KY05110.

\begin{appendix}
\renewcommand{\thetable}{\Roman{table}}

\section{Some expressions and derivations} \label{App}
\subsection{\textup{Coupled eigen equations of the $1^+$ diquark}}\label{App-1}
The four coupled eigen equations for the $1^+$ diquark have the following expressions
\begin{equation}\label{E-Eigen-1+D}
\begin{aligned}
\mu f_3(s)&=+\frac{\mu_1+\mu_2}{e_1+e_2}(e_1+e_2+V^c_1)f_4(s)+\int \frac{\up{d}^3 \vec u}{2e_1e_2}V^c_2\left[ A_{12}(s,u)f_4(u)+A_{14}(s,u)f_6(u) \right],\\
\mu f_4(s)&=+\frac{e_1+e_2}{\mu_1+\mu_2}(e_1+e_2+V^c_1)f_3(s)+\int \frac{\up{d}^3 \vec u}{2e_1e_2}V^c_2\left[ A_{21}(s,u)f_3(u)+A_{23}(s,u)f_5(u) \right],\\
\mu f_5(s)&=-\frac{e_1+e_2}{\mu_1+\mu_2}(e_1+e_2+V^c_1)f_6(s)+\int \frac{\up{d}^3 \vec u}{2e_1e_2}V^c_2\left[ A_{32}(s,u)f_4(u)+A_{34}(s,u)f_6(u) \right],\\
\mu f_6(s)&=-\frac{\mu_1+\mu_2}{e_1+e_2}(e_1+e_2+V^c_1)f_5(s)+\int \frac{\up{d}^3 \vec u}{2e_1e_2}V^c_2\left[ A_{41}(s,u)f_3(u)+A_{43}(s,u)f_5(u) \right],
\end{aligned}
\end{equation}
where $V^c_i=\frac{1}{2}V_{\up{M}i}$; the specific expressions of $A_{ij}(s,u)$ are as
\begin{equation}\notag
\begin{aligned}
%D_1&=\frac{\mu_1+\mu_2}{e_1+e_2}(e_1+e_2+V_\up{Conf1}),~D_3=-D_2,\\
%D_2&=\frac{e_1+e_2}{\mu_1+\mu_2}(e_1+e_2+V_\up{Conf1}),~D_4=-D_1,\\
A_{12}&=\left[ A_-s(e_1-e_2)+\cos\theta(\mu_1e_2+\mu_2e_1) \right]\cos\theta,\quad&
A_{14}&=(\mu_1e_2+\mu_2e_1)(\cos^2\theta-1),\\
A_{21}&=\left[ A_+s(e_1+e_2)+\cos\theta(\mu_1e_2+\mu_2e_1) \right]\cos\theta,\quad&
A_{23}&=(\mu_1e_2+\mu_2e_1)(1-\cos^2\theta),\\
A_{34}&=-\left[ A_+s(e_1+e_2)\cos\theta+\frac{1}{2}(1+\cos^2\theta)(\mu_1e_2+\mu_2e_1) \right],\quad&
A_{32}&=\frac{1}{2}A_{23},\\
A_{43}&=-\left[ A_-s(e_1-e_2)\cos\theta+\frac{1}{2}(1+\cos^2\theta)(\mu_1e_2+\mu_2e_1) \right],\quad&
A_{41}&=\frac{1}{2}A_{14}.
\end{aligned}
\end{equation}
where $\cos\theta=\frac{\vec s\cdot\vec u}{su}$. These four equations can be solved numerically to obtain the diquark mass spectra and corresponding Salpeter wave functions.

\subsection{\textup{Derivations of the baryon Salpeter equation with $1^+$ heavy diquark core}}\label{App-SE}
To reach \eref{E-BSB-D1-2}, firstly, we split out $q_P$ from the propagators, $S(p_2)$ and $D^{\alpha\beta}(p_1)$ then can be expressed as,
\begin{equation}
\begin{gathered}
S(p_2)=-i\left[\frac{\Lambda^+(q_\perp)}{q_P-\zeta_2^+-i\epsilon }+\frac{\Lambda^-(q_\perp)}{q_P-\zeta_2^-+i\epsilon}\right], \\
h_0 D^{\alpha\beta}(p_1)=i \vartheta^{\alpha\beta} \left[\frac{1}{q_P-\zeta_1^++i\epsilon }+\frac{1}{q_P-\zeta_1^--i\epsilon}\right],
\end{gathered}
\end{equation}
where the projector operators are defined as $\Lambda^\pm(p_{2\perp})=\frac{1}{2}\left[1 \pm \hat{H}(p_{2\perp})\right]\gamma^0$; $\zeta^\pm_{1,2}$ are defined as,
\[
\zeta_2^+=\alpha_2 M-\w_2,\quad \zeta_2^-=\alpha_2 M+\w_2,\quad \zeta_1^+=-\alpha_1 M + \w_1, \quad \zeta_1^- = -\alpha_1 M -\w_1.
\]
Performing the integral over $q_P$ on both sides of \eref{E-BS-BD1-1},
%\begin{equation}\notag
%\begin{aligned}
%\varphi^\alpha (q_\perp)
%&=-i\int \frac{\up{d}q_P}{2\pi}B^\alpha(q)\\
%&= -i\int \frac{\up{d}q_P}{2\pi} \left[\frac{\Lambda^+}{q_P-\zeta_2^+-i\epsilon }+\frac{\Lambda^-}{q_P-\zeta_2^-+i\epsilon}\right] \vartheta^{\alpha\beta}\Gamma_\beta (q_\perp) \left[\frac{1}{q_P-\zeta_1^++i\epsilon }+\frac{1}{q_P-\zeta_1^--i\epsilon}\right]\\
%&=\frac{\Lambda^+ \vartheta^{\alpha\beta}\Gamma_\beta (q_\perp)}{\zeta^+_2-\zeta^+_1} + \frac{\Lambda^- \vartheta^{\alpha\beta}\Gamma_\beta (q_\perp) }{\zeta^-_1-\zeta^-_2}
%%&=\frac{\Lambda_1^+ \vartheta^{ab}\Gamma_b (P,q_\perp)}{M-\w_1-\w_2} - \frac{\Lambda_1^- \vartheta^{ab}\Gamma_b (P,q_\perp) }{M+\w_1+\w_2},
%\end{aligned}
%\end{equation}
we obtain the three-dimensional baryon Salpeter equation as
\begin{equation}\label{E-BSB-D1}
\varphi^\alpha (q_\perp)
=\frac{\Lambda^+ \vartheta^{\alpha\beta}\Gamma_\beta (q_\perp)}{M-\w_1-\w_2} - \frac{\Lambda^- \vartheta^{\alpha\beta}\Gamma_\beta (q_\perp) }{M+\w_1+\w_2}.
\end{equation}
By using the projector operators $\Lambda^\pm$, we can get the positive and negative energy Salpeter wave functions as,
\begin{equation}
\begin{aligned}
\varphi^{\alpha+}(q_\perp)&\equiv \Lambda^+   \gamma^0  \varphi^\alpha=+\frac{\Lambda^+ \vartheta^{\alpha\beta}\Gamma_\beta (q_\perp)}{M-\w_1-\w_2},\\
\varphi^{\alpha-}(q_\perp)&\equiv \Lambda^-   \gamma^0  \varphi^\alpha=-\frac{\Lambda^- \vartheta^{\alpha\beta}\Gamma_\beta (q_\perp) }{M+\w_1+\w_2},
%\varphi^\alpha(q_\perp)&=\varphi^{\alpha+}(q_\perp)+\varphi^{\alpha-}(q_\perp),
\end{aligned}
\end{equation}
which are the coupled baryon Salpeter equations with $1^+$ diquark core, and then can be rewritten as the Schr\"{o}dinger-type \eref{E-BSB-D1-2}.

\subsection{\textup{The normalization condition of the baryon Salpeter wave function}}\label{App-2}
To  obtain the normalization of $\varphi_\alpha(q_\perp)$,  we need the inverse of the propagators $D_{\alpha\beta}(p_1)$, which is given by,
\[  D^{-1}_{\alpha\beta}(p_1)=d_{\alpha\beta}D^{-1}(p_1),~~~d^{\alpha\beta}=-g^{\alpha\beta}-\frac{p^\alpha_{1\perp} p^\alpha_{1\perp}}{\w_1^2}, \]
and fulfills
\begin{equation}\notag
D^{-1}_{\alpha\gamma}(p_1) D^{\gamma\beta}(p_1)=d_{\alpha\gamma}\vartheta^{\gamma\beta}=\delta^\beta_\alpha.
 \end{equation}
Note that $d^{\alpha\beta}(p_{1\perp})$ does not explicitly depend on $P^0$.
The BS vertex can also be expressed by the inverse of the propagators as
\[
\Gamma^\alpha(q)%=-i\int \frac{\up{d}^4 k}{(2\pi)^4} V(P,k-q)g^{\alpha\beta}B_{\beta}(P,k)
=S^{-1}(p_2)D^{-1}(p_1)d^{\alpha\beta}B_\beta(q).
\]
The partial differential of the kernel $K^{\alpha\beta}$ gives $\frac{\partial}{\partial P^0} K^{\alpha\beta}(p_1,k_1)= (2\alpha_1)K(k_\perp - q_\perp)g^{\alpha\beta}$.

Finally, putting the these into \eref{E-Norm-B4D} and then calculating the contour integral, we obtain the normalization condition of the Salpeter wave functions as \eref{E-Norm-D1}, where the three-dimensional BS baryon vertex is expressed by the Salpeter wave function as
\begin{equation}
\Gamma^\alpha(P,q_\perp)=\gamma^0 \left[ M\hat{H}(p_{2\perp})-(\w_1+\w_2) \right]d^{\alpha\beta}\varphi_\beta(P,q_\perp).
\end{equation}

%\subsection{\textup{Decomposition of Salpeter wave functions}}

\subsection{\textup{The coupled eigen equations of the $\frac{1}{2}^+$ baryon}}\label{App-3}
By taking different traces on both sides of \eref{E-eigen-D1-1H}, we obtain the following four coupled eigen equations for the $\frac{1}{2}^+$ baryon,
\begin{equation} \label{E-eigen-3H}
\begin{aligned}
Mg_1(\vec q\,)&=D_{1}g_1(\vec q\,)-D_{2}g_2(\vec q\,)+\int \frac{\up d^3 \vec k}{(2\pi)^3} \frac{V_2}{\w_2} \left [  m_2g_1(\vec k) -qcg_2(\vec k)+\frac{1}{2}m_2(c^2-1)g_4(\vec k) \right ],\\
Mg_2(\vec q\,)&=-D_{2}g_1(\vec q\,)-D_{1}g_2(\vec q\,)+\int \frac{\up d^3 \vec k}{(2\pi)^3} \frac{V_2}{\w_2} \left [  -q_1(\vec k) -m_2c g_2(\vec k)-\frac{1}{2} q(c^2-1)g_4(\vec k)\right ],\\
Mg_3(\vec q\,)&=-D_{3}g_1(\vec q\,)-D_{4}g_2(\vec q\,)-D_{5}g_3(\vec q\,) + D_{6} g_4(\vec q\,)\\
         &+ \int \frac{\up d^3 \vec k}{(2\pi)^3} \frac{V_2}{m_1^2\w_2} \left [ -q^3g_1(\vec k) -m_2q^2cg_2(\vec k)-m_2\w_1^2 c g_3(\vec k)+ I_3 g_4(\vec k) \right ], \\
Mg_4(\vec q\,)&=-D_{4}g_1(\vec q\,)+D_{3}g_2(\vec q\,)+D_{6}g_3(\vec q\,) + D_{5} g_4(\vec q\,)\\
         &+ \int \frac{\up d^3 \vec k}{(2\pi)^3} \frac{V_2}{m_1^2\w_2} \left [ -m_2 q^2 g_1(\vec k) +q^3c g_2(\vec k) + q\w_1^2 c g_3(\vec k)+I_4 g_4(\vec k) \right ],
\end{aligned}
\end{equation}
where $V_i=\sigma V_{\up{M}i}$, and $\sigma$, the obtained form factor; $I_3=\frac{1}{2}q(3c^2m_1^2-m_1^2+2c^2q^2)$ and $I_4=\frac{m_2}{q}I_3$; $D_{1}\sim D_6$ have the following explicit expressions
\begin{equation}
\begin{aligned}
D_1&=\frac{m_2}{\w_2}\left( V_1 +\w_1 +\w_2 \right),& D_2&=\frac{q}{m_2}D_1,&D_3&=\frac{q^3V_1}{m_1^2 \w_2},\\
D_5&=\frac{m_2}{\w_2}\left( \frac{\w_1^2}{m_1^2}V_1+\w_1+\w_2  \right),&D_4&=\frac{m_2}{q}D_3,&D_6&=\frac{q}{m_2}D_5.
\end{aligned}
\end{equation}

\subsection{\textup{Spin-weighted average $V_0$ in diquark-quark interaction}}\label{App-5}
 The baryon under different diquark basis has the following relationships,
\begin{equation}
\begin{bmatrix} \ket{(12)_03} \\ \ket{(12)_13} \end{bmatrix}=
\begin{bmatrix} -\frac{1}{2}& -\frac{\sqrt{3}}{2} \\ +\frac{\sqrt{3}}{2} & -\frac{1}{2}\end{bmatrix}
\begin{bmatrix} \ket{1(23)_0} \\ \ket{1(23)_1} \end{bmatrix}=
\begin{bmatrix} -\frac{1}{2}& +\frac{\sqrt{3}}{2} \\ -\frac{\sqrt{3}}{2} & -\frac{1}{2}\end{bmatrix}
\begin{bmatrix} \ket{(31)_02} \\ \ket{(31)_12} \end{bmatrix},
\end{equation}
where $\ket{(12)_03}$ denotes the baryon state when the quark-1 and quark-2 inside the baryon form the spin-0 diquark, and then others are implied. Note that the above relations can be considered a counter-clock rotation within different diquark basis and the rotation angles are respectively $120^\circ$ and $-120^\circ$.

Now take the $\Xi^{++}_{cc}$ as an example to show how the corresponding $V_0$ value are decided. In $\Xi^{++}_{cc}$, above equation implies $\ket{(cc)_1u} = \frac{\sqrt{3}}{2} \ket{c(cu)_0} -\frac{1}{2}\ket{c(cu)_1}$ and then the $V_0$ between the $(cc)_1$-diquark and the $u$-quark are determined by $V_0[(cu)_1]$ and $V_0[(cu)_0]$ which corresponds to that of the mesons $D^{*0}$ and $D^0$ respectively. Then  considering above relations, we can express the $V_0(\Xi^{++}_{cc})$ as
\[  V_0(\Xi^{++}_{cc})=0.75V_0(D^0)+ 0.25V_0(D^{*0}), \]
where the involved $V_0$ of the mesons are listed in \autoref{Tab-V0-mesons}.
\begin{table}[h!]
\caption{The relevant parameters $(-V_0)$ in GeV for the $0^-$ and $1^-$ mesons.}
\label{Tab-V0-mesons}
\vspace{0.2em}\centering
\begin{tabular}{ c|ccccccccccccc }
\toprule[2pt]
$Q\bar q$ 	  & $c\bar u$		& $c\bar d$		& $c\bar s$	&$b\bar u$	& $b\bar s$\\
\midrule[1.5pt]
$J^P=0^-$  	  & $0.512$	  	&$0.509$	  		&$0.489$	  	&$0.341$	  	&$0.322$	  \\
$J^P=1^-$	  & $0.378$	  	&$0.376$	  		&$0.352$	  	&$0.296$	  	&$0.275$	\\
\bottomrule[2pt]
\end{tabular}
\end{table}
which are obtained by solving the corresponding meson BSE, namely, \,(\ref{E-BS-HM2}).
The obtained $V_0$ for the doubly heavy baryons are listed in \autoref{Tab-V0}. Note that now all the involved model parameters in this work are determined by the corresponding meson spectra.
\end{appendix}
\medskip

\biboptions{numbers,sort&compress}
\setlength{\bibsep}{0.5ex}  % vertical spacing between references
%\bibliography{../reference-QIANG}

\begin{thebibliography}{10}
\expandafter\ifx\csname url\endcsname\relax
  \def\url#1{\texttt{#1}}\fi
\expandafter\ifx\csname urlprefix\endcsname\relax\def\urlprefix{URL }\fi
\expandafter\ifx\csname href\endcsname\relax
  \def\href#1#2{#2} \def\path#1{#1}\fi

\bibitem{LHCb2017B}
R.~Aaij, et~al., Phys. Rev. Lett. 119~(11) (2017) 112001.
\newblock \href {http://arxiv.org/abs/1707.01621} {\path{arXiv:1707.01621}},
  \href {https://doi.org/10.1103/PhysRevLett.119.112001}
  {\path{DOI:10.1103/PhysRevLett.119.112001}}.

\bibitem{LHCb2018}
R.~Aaij, et~al., Phys. Rev. Lett. 121~(5) (2018) 052002.
\newblock \href {http://arxiv.org/abs/1806.02744} {\path{arXiv:1806.02744}},
  \href {https://doi.org/10.1103/PhysRevLett.121.052002}
  {\path{DOI:10.1103/PhysRevLett.121.052002}}.

\bibitem{LHCb2017A}
R.~Aaij, et~al., Phys. Rev. Lett. 118~(18) (2017) 182001.
\newblock \href {http://arxiv.org/abs/1703.04639} {\path{arXiv:1703.04639}},
  \href {https://doi.org/10.1103/PhysRevLett.118.182001}
  {\path{DOI:10.1103/PhysRevLett.118.182001}}.

\bibitem{LHCb2018-Xib}
R.~Aaij, et~al., Phys. Rev. Lett. 121~(7) (2018) 072002.
\newblock \href {http://arxiv.org/abs/1805.09418} {\path{arXiv:1805.09418}},
  \href {https://doi.org/10.1103/PhysRevLett.121.072002}
  {\path{DOI:10.1103/PhysRevLett.121.072002}}.

\bibitem{Chernyshev1996}
S.~Chernyshev, M.~A. Nowak, I.~Zahed, Phys. Rev. D 53 (1996) 5176--5184.
\newblock \href {http://arxiv.org/abs/hep-ph/9510326}
  {\path{arXiv:hep-ph/9510326}}, \href
  {https://doi.org/10.1103/PhysRevD.53.5176}
  {\path{DOI:10.1103/PhysRevD.53.5176}}.

\bibitem{Gershtein1998}
S.~S. Gershtein, V.~V. Kiselev, A.~K. Likhoded, A.~I. Onishchenko, Acta Phys.
  Hung. A 9 (1999) 133--144.
\newblock \href {http://arxiv.org/abs/hep-ph/9811212}
  {\path{arXiv:hep-ph/9811212}}, \href {https://doi.org/10.1134/1.855633}
  {\path{DOI:10.1134/1.855633}}.

\bibitem{Gershtein1999}
S.~S. Gershtein, V.~V. Kiselev, A.~K. Likhoded, A.~I. Onishchenko, Mod. Phys.
  Lett. A 14 (1999) 135--146.
\newblock \href {http://arxiv.org/abs/hep-ph/9807375}
  {\path{arXiv:hep-ph/9807375}}, \href
  {https://doi.org/10.1142/S0217732399000171}
  {\path{DOI:10.1142/S0217732399000171}}.

\bibitem{Gershtein2000}
S.~S. Gershtein, V.~V. Kiselev, A.~K. Likhoded, A.~I. Onishchenko, Phys. Rev. D
  62 (2000) 054021.
\newblock \href {https://doi.org/10.1103/PhysRevD.62.054021}
  {\path{DOI:10.1103/PhysRevD.62.054021}}.

\bibitem{Ebert2002}
D.~Ebert, R.~N. Faustov, V.~O. Galkin, A.~P. Martynenko, Phys. Rev. D 66 (2002)
  014008.
\newblock \href {http://arxiv.org/abs/hep-ph/0201217}
  {\path{arXiv:hep-ph/0201217}}, \href
  {https://doi.org/10.1103/PhysRevD.66.014008}
  {\path{DOI:10.1103/PhysRevD.66.014008}}.

\bibitem{HeDH2004}
D.-H. He, K.~Qian, Y.-B. Ding, X.-Q. Li, P.-N. Shen, Phys. Rev. D70 (2004)
  094004.
\newblock \href {http://arxiv.org/abs/hep-ph/0403301}
  {\path{arXiv:hep-ph/0403301}}, \href
  {https://doi.org/10.1103/PhysRevD.70.094004}
  {\path{DOI:10.1103/PhysRevD.70.094004}}.

\bibitem{Chang2006}
C.-H. Chang, C.-F. Qiao, J.-X. Wang, X.-G. Wu, Phys. Rev. D 73 (2006) 094022.
\newblock \href {http://arxiv.org/abs/hep-ph/0601032}
  {\path{arXiv:hep-ph/0601032}}, \href
  {https://doi.org/10.1103/PhysRevD.73.094022}
  {\path{DOI:10.1103/PhysRevD.73.094022}}.

\bibitem{ZhangJR2008}
J.-R. Zhang, M.-Q. Huang, Phys. Rev. D 78 (2008) 094007.
\newblock \href {http://arxiv.org/abs/0810.5396} {\path{arXiv:0810.5396}},
  \href {https://doi.org/10.1103/PhysRevD.78.094007}
  {\path{DOI:10.1103/PhysRevD.78.094007}}.

\bibitem{WangZG2010}
Z.-G. Wang, Eur. Phys. J. A45 (2010) 267--274.
\newblock \href {http://arxiv.org/abs/1001.4693} {\path{arXiv:1001.4693}},
  \href {https://doi.org/10.1140/epja/i2010-11004-3}
  {\path{DOI:10.1140/epja/i2010-11004-3}}.

\bibitem{Brodsky2011}
S.~J. Brodsky, F.-K. Guo, C.~Hanhart, U.-G. Meissner, Phys. Lett. B 698 (2011)
  251--255.
\newblock \href {http://arxiv.org/abs/1101.1983} {\path{arXiv:1101.1983}},
  \href {https://doi.org/10.1016/j.physletb.2011.03.014}
  {\path{DOI:10.1016/j.physletb.2011.03.014}}.

\bibitem{Aliev2012}
T.~M. Aliev, K.~Azizi, M.~Savci, Nucl. Phys. A 895 (2012) 59--70.
\newblock \href {http://arxiv.org/abs/1205.2873} {\path{arXiv:1205.2873}},
  \href {https://doi.org/10.1016/j.nuclphysa.2012.09.009}
  {\path{DOI:10.1016/j.nuclphysa.2012.09.009}}.

\bibitem{SunZF2015}
Z.-F. Sun, Z.-W. Liu, X.~Liu, S.-L. Zhu, Phys. Rev. D91~(9) (2015) 094030.
\newblock \href {http://arxiv.org/abs/1411.2117} {\path{arXiv:1411.2117}},
  \href {https://doi.org/10.1103/PhysRevD.91.094030}
  {\path{DOI:10.1103/PhysRevD.91.094030}}.

\bibitem{WeiKW2015}
K.-W. Wei, B.~Chen, X.-H. Guo, Phys. Rev. D 92~(7) (2015) 076008.
\newblock \href {http://arxiv.org/abs/1503.05184} {\path{arXiv:1503.05184}},
  \href {https://doi.org/10.1103/PhysRevD.92.076008}
  {\path{DOI:10.1103/PhysRevD.92.076008}}.

\bibitem{SunZF2016}
Z.-F. Sun, M.~J. Vicente~Vacas, Phys. Rev. D 93~(9) (2016) 094002.
\newblock \href {http://arxiv.org/abs/1602.04714} {\path{arXiv:1602.04714}},
  \href {https://doi.org/10.1103/PhysRevD.93.094002}
  {\path{DOI:10.1103/PhysRevD.93.094002}}.

\bibitem{Lewis2001}
R.~Lewis, N.~Mathur, R.~M. Woloshyn, Phys. Rev. D64 (2001) 094509.
\newblock \href {http://arxiv.org/abs/hep-ph/0107037}
  {\path{arXiv:hep-ph/0107037}}, \href
  {https://doi.org/10.1103/PhysRevD.64.094509}
  {\path{DOI:10.1103/PhysRevD.64.094509}}.

\bibitem{Flynn2003}
J.~M. Flynn, F.~Mescia, A.~S.~B. Tariq, JHEP 07 (2003) 066.
\newblock \href {http://arxiv.org/abs/hep-lat/0307025}
  {\path{arXiv:hep-lat/0307025}}, \href
  {https://doi.org/10.1088/1126-6708/2003/07/066}
  {\path{DOI:10.1088/1126-6708/2003/07/066}}.

\bibitem{Brown2014}
Z.~S. Brown, W.~Detmold, S.~Meinel, K.~Orginos, Phys. Rev. D 90~(9) (2014)
  094507.
\newblock \href {http://arxiv.org/abs/1409.0497} {\path{arXiv:1409.0497}},
  \href {https://doi.org/10.1103/PhysRevD.90.094507}
  {\path{DOI:10.1103/PhysRevD.90.094507}}.

\bibitem{Karliner2014}
M.~Karliner, J.~L. Rosner, Phys. Rev. D 90~(9) (2014) 094007.
\newblock \href {http://arxiv.org/abs/1408.5877} {\path{arXiv:1408.5877}},
  \href {https://doi.org/10.1103/PhysRevD.90.094007}
  {\path{DOI:10.1103/PhysRevD.90.094007}}.

\bibitem{Kiselev1999}
V.~V. Kiselev, A.~K. Likhoded, A.~I. Onishchenko, Phys. Rev. D 60 (1999)
  014007.
\newblock \href {http://arxiv.org/abs/hep-ph/9807354}
  {\path{arXiv:hep-ph/9807354}}, \href
  {https://doi.org/10.1103/PhysRevD.60.014007}
  {\path{DOI:10.1103/PhysRevD.60.014007}}.

\bibitem{Chang2008}
C.-H. Chang, T.~Li, X.-Q. Li, Y.-M. Wang, Commun. Theor. Phys. 49 (2008)
  993--1000.
\newblock \href {http://arxiv.org/abs/0704.0016} {\path{arXiv:0704.0016}},
  \href {https://doi.org/10.1088/0253-6102/49/4/38}
  {\path{DOI:10.1088/0253-6102/49/4/38}}.

\bibitem{Berezhnoy2016}
A.~V. Berezhnoy, A.~K. Likhoded, Phys. Atom. Nucl. 79~(2) (2016) 260--265.
\newblock \href {https://doi.org/10.1134/S1063778816010087}
  {\path{DOI:10.1134/S1063778816010087}}.

\bibitem{Cahill1987}
R.~T. Cahill, C.~D. Roberts, J.~Praschifka, Phys. Rev. D 36 (1987) 2804.
\newblock \href {https://doi.org/10.1103/PhysRevD.36.2804}
  {\path{DOI:10.1103/PhysRevD.36.2804}}.

\bibitem{Keiner1996}
V.~Keiner, Z. Phys. A 354 (1996) 87.
\newblock \href {http://arxiv.org/abs/hep-ph/9509284}
  {\path{arXiv:hep-ph/9509284}}, \href {https://doi.org/10.1007/s002180050015}
  {\path{DOI:10.1007/s002180050015}}.

\bibitem{Keiner1996A}
V.~Keiner, Phys. Rev. C 54 (1996) 3232--3239.
\newblock \href {http://arxiv.org/abs/hep-ph/9603226}
  {\path{arXiv:hep-ph/9603226}}, \href
  {https://doi.org/10.1103/PhysRevC.54.3232}
  {\path{DOI:10.1103/PhysRevC.54.3232}}.

\bibitem{Maris2002}
P.~Maris, Few Body Syst. 32 (2002) 41--52.
\newblock \href {http://arxiv.org/abs/nucl-th/0204020}
  {\path{arXiv:nucl-th/0204020}}, \href
  {https://doi.org/10.1007/s00601-002-0111-7}
  {\path{DOI:10.1007/s00601-002-0111-7}}.

\bibitem{Maris2004}
P.~Maris, Few Body Syst. 35 (2004) 117--127.
\newblock \href {http://arxiv.org/abs/nucl-th/0409008}
  {\path{arXiv:nucl-th/0409008}}, \href
  {https://doi.org/10.1007/s00601-004-0064-0}
  {\path{DOI:10.1007/s00601-004-0064-0}}.

\bibitem{Maris2005}
P.~Maris, AIP Conf. Proc. 768 (2005) 256--258.
\newblock \href {http://arxiv.org/abs/nucl-th/0412059}
  {\path{arXiv:nucl-th/0412059}}, \href {https://doi.org/10.1063/1.1932926}
  {\path{DOI:10.1063/1.1932926}}.

\bibitem{GuoXH1999}
X.-H. Guo, A.~W. Thomas, A.~G. Williams, Phys. Rev. D 59 (1999) 116007.
\newblock \href {http://arxiv.org/abs/hep-ph/9805331}
  {\path{arXiv:hep-ph/9805331}}, \href
  {https://doi.org/10.1103/PhysRevD.59.116007}
  {\path{DOI:10.1103/PhysRevD.59.116007}}.

\bibitem{GuoXH2007}
X.-H. Guo, H.-K. Wu, Phys. Lett. B 654 (2007) 97--103.
\newblock \href {http://arxiv.org/abs/0705.1379} {\path{arXiv:0705.1379}},
  \href {https://doi.org/10.1016/j.physletb.2007.05.007}
  {\path{DOI:10.1016/j.physletb.2007.05.007}}.

\bibitem{WengMH2011}
M.~H. Weng, X.~H. Guo, A.~W. Thomas, Phys. Rev. D 83 (2011) 056006.
\newblock \href {http://arxiv.org/abs/1012.0082} {\path{arXiv:1012.0082}},
  \href {https://doi.org/10.1103/PhysRevD.83.056006}
  {\path{DOI:10.1103/PhysRevD.83.056006}}.

\bibitem{ZhangL2013}
L.~Zhang, X.~H. Guo, Phys. Rev. D 87~(7) (2013) 076013.
\newblock \href {http://arxiv.org/abs/1305.1078} {\path{arXiv:1305.1078}},
  \href {https://doi.org/10.1103/PhysRevD.87.076013}
  {\path{DOI:10.1103/PhysRevD.87.076013}}.

\bibitem{LiuY2015}
Y.~Liu, X.~H. Guo, C.~Wang, Phys. Rev. D91~(1) (2015) 016006.
\newblock \href {https://doi.org/10.1103/PhysRevD.91.016006}
  {\path{DOI:10.1103/PhysRevD.91.016006}}.

\bibitem{WeiKW2017}
K.-W. Wei, B.~Chen, N.~Liu, Q.-Q. Wang, X.-H. Guo, Phys. Rev. D 95~(11) (2017)
  116005.
\newblock \href {http://arxiv.org/abs/1609.02512} {\path{arXiv:1609.02512}},
  \href {https://doi.org/10.1103/PhysRevD.95.116005}
  {\path{DOI:10.1103/PhysRevD.95.116005}}.

\bibitem{LiuLL2017}
L.-L. Liu, C.~Wang, Y.~Liu, X.-H. Guo, Phys. Rev. D 95~(5) (2017) 054001.
\newblock \href {http://arxiv.org/abs/1612.06084} {\path{arXiv:1612.06084}},
  \href {https://doi.org/10.1103/PhysRevD.95.054001}
  {\path{DOI:10.1103/PhysRevD.95.054001}}.

\bibitem{YUQX2018}
Q.-X. Yu, X.-H. Guo (2018).
\newblock \href {http://arxiv.org/abs/1810.00437} {\path{arXiv:1810.00437}}.

\bibitem{Chang2005A}
C.-H. Chang, J.-K. Chen, X.-Q. Li, G.-L. Wang, Commun. Theor. Phys. 43 (2005)
  113--118.
\newblock \href {http://arxiv.org/abs/hep-ph/0406050}
  {\path{arXiv:hep-ph/0406050}}, \href
  {https://doi.org/10.1088/0253-6102/43/1/023}
  {\path{DOI:10.1088/0253-6102/43/1/023}}.

\bibitem{Chang2010}
C.~H. Chang, G.~L. Wang, Sci. China Phys. Mech. Astron. 53 (2010) 2005--2018.
\newblock \href {http://arxiv.org/abs/1003.3827} {\path{arXiv:1003.3827}},
  \href {https://doi.org/10.1007/s11433-010-4156-1}
  {\path{DOI:10.1007/s11433-010-4156-1}}.

\bibitem{Chang2005}
C.-H. Chang, C.~Kim, G.-L. Wang, Phys. Lett. B 623 (2005) 218--226.
\newblock \href {https://doi.org/10.1016/j.physletb.2005.07.059}
  {\path{DOI:10.1016/j.physletb.2005.07.059}}.

\bibitem{WangZ2012A}
Z.-H. Wang, G.-L. Wang, C.-H. Chang, J. Phys. G: Nucl. Part. Phys. 39 (2012)
  015009.
\newblock \href {http://arxiv.org/abs/1107.0474} {\path{arXiv:1107.0474}},
  \href {https://doi.org/10.1088/0954-3899/39/1/015009}
  {\path{DOI:10.1088/0954-3899/39/1/015009}}.

\bibitem{WangT2013}
T.~Wang, G.-L. Wang, H.-F. Fu, W.-L. Ju, JHEP 07 (2013) 120.
\newblock \href {http://arxiv.org/abs/1305.1067} {\path{arXiv:1305.1067}},
  \href {https://doi.org/10.1007/JHEP07(2013)120}
  {\path{DOI:10.1007/JHEP07(2013)120}}.

\bibitem{WangT2013A}
T.~Wang, G.-L. Wang, W.-L. Ju, Y.~Jiang, JHEP 03 (2013) 110.
\newblock \href {http://arxiv.org/abs/1303.1563} {\path{arXiv:1303.1563}},
  \href {https://doi.org/10.1007/JHEP03(2013)110}
  {\path{DOI:10.1007/JHEP03(2013)110}}.

\bibitem{LiQ2016}
Q.~Li, T.~Wang, Y.~Jiang, H.~Yuan, G.-L. Wang, Eur. Phys. J. C 76~(8) (2016)
  454.
\newblock \href {https://doi.org/10.1140/epjc/s10052-016-4306-3}
  {\path{DOI:10.1140/epjc/s10052-016-4306-3}}.

\bibitem{LiQ2017}
Q.~Li, T.~Wang, Y.~Jiang, H.~Yuan, T.~Zhou, G.-L. Wang, Eur. Phys. J. C 77~(1)
  (2017) 12.
\newblock \href {https://doi.org/10.1140/epjc/s10052-016-4588-5}
  {\path{DOI:10.1140/epjc/s10052-016-4588-5}}.

\bibitem{LiQ2017A}
Q.~Li, Y.~Jiang, T.~Wang, H.~Yuan, G.-L. Wang, C.-H. Chang, Eur. Phys. J. C
  77~(5) (2017) 297.
\newblock \href {http://arxiv.org/abs/1701.03252} {\path{arXiv:1701.03252}},
  \href {https://doi.org/10.1140/epjc/s10052-017-4865-y}
  {\path{DOI:10.1140/epjc/s10052-017-4865-y}}.

\bibitem{Kim2004}
C.~S. Kim, G.-L. Wang, Phys. Lett. B 584 (2004) 285--293.
\newblock \href {http://arxiv.org/abs/hep-ph/0309162}
  {\path{arXiv:hep-ph/0309162}}, \href
  {https://doi.org/10.1016/j.physletb.2004.01.058}
  {\path{DOI:10.1016/j.physletb.2004.01.058}}.

\bibitem{Eichten1978}
E.~Eichten, K.~Gottfried, T.~Kinoshita, K.~D. Lane, T.-M. Yan, Phys. Rev. D 17
  (1978) 3090.
\newblock \href {https://doi.org/10.1103/PhysRevD.17.3090,
  10.1103/physrevd.21.313.2} {\path{DOI:10.1103/PhysRevD.17.3090,
  10.1103/physrevd.21.313.2}}.

\bibitem{Eichten1980}
E.~Eichten, K.~Gottfried, T.~Kinoshita, K.~D. Lane, T.-M. Yan, Phys. Rev. D 21
  (1980) 203.
\newblock \href {https://doi.org/10.1103/PhysRevD.21.203}
  {\path{DOI:10.1103/PhysRevD.21.203}}.

\bibitem{Laermann1986}
E.~Laermann, F.~Langhammer, I.~Schmitt, P.~M. Zerwas, Phys. Lett. B 173 (1986)
  437--442.
\newblock \href {https://doi.org/10.1016/0370-2693(86)90411-9}
  {\path{DOI:10.1016/0370-2693(86)90411-9}}.

\bibitem{Born1989}
K.~D. Born, E.~Laermann, N.~Pirch, T.~F. Walsh, P.~M. Zerwas, Phys. Rev. D 40
  (1989) 1653--1663.
\newblock \href {https://doi.org/10.1103/PhysRevD.40.1653}
  {\path{DOI:10.1103/PhysRevD.40.1653}}.

\bibitem{Chao1992}
K.-T. Chao, Y.-B. Ding, D.-H. Qin, Commun. Theor. Phys. 18 (1992) 321--326.

\bibitem{DingYB1993}
Y.-B. Ding, K.-T. Chao, D.-H. Qin, Chin. Phys. Lett. 10 (1993) 460--463.
\newblock \href {https://doi.org/10.1088/0256-307X/10/8/004}
  {\path{DOI:10.1088/0256-307X/10/8/004}}.

\bibitem{DingYB1995}
Y.-B. Ding, K.-T. Chao, D.-H. Qin, Phys. Rev. D 51 (1995) 5064--5068.
\newblock \href {http://arxiv.org/abs/hep-ph/9502409}
  {\path{arXiv:hep-ph/9502409}}, \href
  {https://doi.org/10.1103/PhysRevD.51.5064}
  {\path{DOI:10.1103/PhysRevD.51.5064}}.

\bibitem{Munz1994}
C.~R. Munz, J.~Resag, B.~C. Metsch, H.~R. Petry, Nucl. Phys. A 578 (1994)
  418--440.
\newblock \href {http://arxiv.org/abs/nucl-th/9307027}
  {\path{arXiv:nucl-th/9307027}}, \href
  {https://doi.org/10.1016/0375-9474(94)90754-4}
  {\path{DOI:10.1016/0375-9474(94)90754-4}}.

\bibitem{Resag1995}
J.~Resag, C.~R. Munz, Nucl. Phys. A 590 (1995) 735.
\newblock \href {http://arxiv.org/abs/nucl-th/9407033}
  {\path{arXiv:nucl-th/9407033}}, \href
  {https://doi.org/10.1016/0375-9474(95)00203-D}
  {\path{DOI:10.1016/0375-9474(95)00203-D}}.

\bibitem{Parramore1995}
J.~Parramore, J.~Piekarewicz, Nucl. Phys. A 585 (1995) 705--726.
\newblock \href {http://arxiv.org/abs/nucl-th/9402019}
  {\path{arXiv:nucl-th/9402019}}, \href
  {https://doi.org/10.1016/0375-9474(94)00722-Y}
  {\path{DOI:10.1016/0375-9474(94)00722-Y}}.

\bibitem{Zoller1995}
G.~Zoller, S.~Hainzl, C.~R. Munz, M.~Beyer, Z. Phys. C 68 (1995) 103--112.
\newblock \href {http://arxiv.org/abs/hep-ph/9412355}
  {\path{arXiv:hep-ph/9412355}}, \href {https://doi.org/10.1007/BF01579809}
  {\path{DOI:10.1007/BF01579809}}.

\bibitem{Babutsidze1998}
T.~Babutsidze, T.~Kopaleishvili, A.~Rusetsky, Phys. Lett. B 426 (1998)
  139--148.
\newblock \href {http://arxiv.org/abs/hep-ph/9710278}
  {\path{arXiv:hep-ph/9710278}}, \href
  {https://doi.org/10.1016/S0370-2693(98)00259-7}
  {\path{DOI:10.1016/S0370-2693(98)00259-7}}.

\bibitem{Karliner2017}
M.~Karliner, J.~L. Rosner, Nature 551 (2017) 89.
\newblock \href {http://arxiv.org/abs/1708.02547} {\path{arXiv:1708.02547}},
  \href {https://doi.org/10.1038/nature24289} {\path{DOI:10.1038/nature24289}}.

\bibitem{Olsson1995}
M.~G. Olsson, S.~Veseli, K.~Williams, Phys. Rev. D 52 (1995) 5141--5151.
\newblock \href {http://arxiv.org/abs/hep-ph/9503477}
  {\path{arXiv:hep-ph/9503477}}, \href
  {https://doi.org/10.1103/PhysRevD.52.5141}
  {\path{DOI:10.1103/PhysRevD.52.5141}}.

\bibitem{Salpeter1952}
E.~E. Salpeter, Phys. Rev. 87 (1952) 328--343.
\newblock \href {https://doi.org/10.1103/PhysRev.87.328}
  {\path{DOI:10.1103/PhysRev.87.328}}.

\bibitem{Behrends1957}
R.~E. Behrends, C.~Fronsdal, Phys. Rev. 106~(2) (1957) 345.
\newblock \href {https://doi.org/10.1103/PhysRev.106.345}
  {\path{DOI:10.1103/PhysRev.106.345}}.

\bibitem{Mathur2018}
N.~Mathur, M.~Padmanath, S.~Mondal, Phys. Rev. Lett. 121~(20) (2018) 202002.
\newblock \href {http://arxiv.org/abs/1806.04151} {\path{arXiv:1806.04151}},
  \href {https://doi.org/10.1103/PhysRevLett.121.202002}
  {\path{DOI:10.1103/PhysRevLett.121.202002}}.

\bibitem{Mathur2019}
N.~Mathur, M.~Padmanath, Phys. Rev. D99~(3) (2019) 031501.
\newblock \href {http://arxiv.org/abs/1807.00174} {\path{arXiv:1807.00174}},
  \href {https://doi.org/10.1103/PhysRevD.99.031501}
  {\path{DOI:10.1103/PhysRevD.99.031501}}.

\bibitem{Karliner2018}
M.~Karliner, J.~L. Rosner, Phys. Rev. D 97~(9) (2018) 094006.
\newblock \href {http://arxiv.org/abs/1803.01657} {\path{arXiv:1803.01657}},
  \href {https://doi.org/10.1103/PhysRevD.97.094006}
  {\path{DOI:10.1103/PhysRevD.97.094006}}.

\bibitem{Albertus2007}
C.~Albertus, E.~Hernandez, J.~Nieves, J.~M. Verde-Velasco, Eur. Phys. J. A 32
  (2007) 183--199.
\newblock \href {http://arxiv.org/abs/hep-ph/0610030}
  {\path{arXiv:hep-ph/0610030}}, \href
  {https://doi.org/10.1140/epja/i2007-10364-y}
  {\path{DOI:10.1140/epja/i2007-10364-y}}.

\bibitem{Giannuzzi2009}
F.~Giannuzzi, Phys. Rev. D 79 (2009) 094002.
\newblock \href {http://arxiv.org/abs/0902.4624} {\path{arXiv:0902.4624}},
  \href {https://doi.org/10.1103/PhysRevD.79.094002}
  {\path{DOI:10.1103/PhysRevD.79.094002}}.

\bibitem{WengXZ2018}
X.-Z. Weng, X.-L. Chen, W.-Z. Deng, Phys. Rev. D 97~(5) (2018) 054008.
\newblock \href {http://arxiv.org/abs/1801.08644} {\path{arXiv:1801.08644}},
  \href {https://doi.org/10.1103/PhysRevD.97.054008}
  {\path{DOI:10.1103/PhysRevD.97.054008}}.

\bibitem{LvQF2017}
Q.-F. L\"u, K.-L. Wang, L.-Y. Xiao, X.-H. Zhong, Phys. Rev. D96~(11) (2017)
  114006.
\newblock \href {http://arxiv.org/abs/1708.04468} {\path{arXiv:1708.04468}},
  \href {https://doi.org/10.1103/PhysRevD.96.114006}
  {\path{DOI:10.1103/PhysRevD.96.114006}}.

\end{thebibliography}

\end{document}